\documentclass[prb,twocolumn,floatfix,showpacs]{revtex4-1}
\usepackage{amsmath,amssymb,graphicx,psfrag}
\usepackage{color}
\usepackage{multirow}
\def\bea{\begin{eqnarray}}
\def\eea{\end{eqnarray}}
\newcommand{\rv}{\vec{r}}

\newcommand{\kv}{{\vec{k}}}

\newcommand{\kvj}{\vec{k}^{(j)}}

\newcommand{\uj}{u^{(j)}}

\newcommand{\ujs}{u^{(j)*}}
\newcommand{\duj}{\dot{u}^{(j)}}
\newcommand{\dduj}{\ddot{u}^{(j)}}
\newcommand{\dujs}{\dot{u}^{(j)*}}
\newcommand{\ddujs}{\ddot{u}^{(j)*}}
\newcommand{\dujp}{\dot{u}^{(j)'}}

\newcommand{\kh}{{\hat{k}}}

\newcommand{\khj}{\hat{k}^{(j)}}
\newcommand{\Rv}{\vec{R}}

\newcommand{\pj}{p^{(j)}}
\newcommand{\pjs}{p^{(j)*}}
\newcommand{\rj}{r^{(j)}}
\newcommand{\rjs}{r^{(j)*}}
\newcommand{\drj}{\dot{r}^{(j)}}
\newcommand{\drjs}{\dot{r}^{(j)*}}
\newcommand{\cjam}{c_{j, a}^{-}}
\newcommand{\cjams}{c_{j, a}^{-*}}
\newcommand{\cjamn}{c_{j, a}^{-,0}}
\newcommand{\cjamns}{c_{j, a}^{-,0*}}
\newcommand{\cjap}{c_{j, a}^{+}}
\newcommand{\cjapn}{c_{j, a}^{+,0}}
\newcommand{\cjaps}{c_{j, a}^{+*}}
\newcommand{\cjbm}{c_{j, b}^{-}}
\newcommand{\cjbmn}{c_{j, b}^{-,0}}
\newcommand{\cjbmns}{c_{j, b}^{-,0*}}
\newcommand{\cjbms}{c_{j, b}^{-*}}
\newcommand{\cjbp}{c_{j, b}^{+}}
\newcommand{\cjbpn}{c_{j, b}^{+,0}}
\newcommand{\cjbps}{c_{j, b}^{+*}}
\newcommand{\nh}{\hat{n}}
\newcommand{\zh}{\hat{z}}
\newcommand{\xh}{\hat{x}}
\newcommand{\yh}{\hat{y}}

\newcommand{\Aj}{A^{(j)}}
\newcommand{\Ajs}{A^{(j)*}}
\newcommand{\epsnew}{\tilde{\epsilon}}
\newcommand{\kvjd}{\vec{k}^{(j)\dagger}}
\newcommand{\rotM}{\mathbf{M}}
\newcommand{\rotR}{\mathbf{R}}
\newcommand{\rotI}{\mathbf{I}}

\begin{document}

\title{Structural short-range forces between solid-melt interfaces}

\date{\today}

\author{R. Spatschek}
\affiliation{Max-Planck-Institut f\"ur Eisenforschung GmbH, D-40237 D\"usseldorf, Germany}
\author{A. Adland}
\affiliation{Physics Department and Center for Interdisciplinary Research on Complex Systems, Northeastern University, Boston, MA 02115, USA}
\author{A. Karma}
\affiliation{Physics Department and Center for Interdisciplinary Research on Complex Systems, Northeastern University, Boston, MA 02115, USA}

\begin{abstract}
We predict the structural interaction of crystalline solid-melt interfaces using amplitude equations which are derived from classical density functional theory or phase-field-crystal modeling.
The solid ordering decays exponentially on the scale of the interface thickness at solid-melt interfaces;
the overlap of two such profiles leads to a short range interaction, which is mainly carried by the longest-range density waves, which can facilitate grain boundary premelting.
We calculate the tail of these interactions, depending on the relative translation of the two crystals fully analytically and predict the interaction potential, and compare it to numerical simulations.
For grain boundaries the interaction is predicted to decay twice faster as for two crystals without misorientation.
\end{abstract}


\maketitle 

\section{Introduction}

Grain boundaries (GBs) and interfaces in general have a strong influence on mechanical behavior and other materials properties. 
Therefore, they have been widely studied both experimentally \cite{SuttonBalluffi1995} and computationally \cite{Mishinetal2010} for a long time. 
At high temperatures close to the melting point, GBs can display pronounced disorder, even leading to the formation of nanometer-scale intergranular films with liquid-like properties. 
The formation of those films below the bulk melting point, known as GB premelting, lead to catastrophic materials failure, initiated by an enormous reduction of the shear resistance.
This phenomenon is of interest for predicting the formation of solidification defects associated with the formation of those intergranular films, which can lead to hot cracking during the late stages of solidification \cite{Rappazetal2003,Wangetal2004,Astaetal2009}, and more generally for understanding the microstructure and mechanical behavior of structural alloys at high homologous temperature.

Over the years, there have been many experimental \cite{Chanetal1985,BallufiMaurer1988,HsiehBalluffi1989,Masumuraetal1972,Voldetal1972,Watanabeetal1984,Dashetal1995,Inokoetal1997,Divinskietal2005,Luoetal2005,Guptaetal2007} and theoretical investigations of GB premelting.
The latter include discrete lattice models \cite{KikuchiCahn1980} and molecular dynamics (MD) simulations \cite{BroughtonGilmer1986,vonAlfthan2007,WilliamsMishin2009,Hoytetal2009,Fensinetal2010,Olmstedetal2011}, as well as conventional phase-field models \cite{LobkovskyWarren2002,Tangetal2006,Mishinetal2009,Wangetal2008}, which either exploit an orientational order parameter \cite{LobkovskyWarren2002,Tangetal2006} or multiple phase-fields \cite{Mishinetal2009,Wangetal2008} to distinguish between grains, and the phase-field-crystal (PFC) method \cite{Berryetal2008,Mellenthinetal2008}, which resolves the crystal density field on an atomic scale and hence naturally models crystal defects such as isolated dislocations and GBs. 

The determination of the premelted layer width $W$ and the quantification of the fundamental forces that control this width are of striking interest for GB premelting.
Experimentally, these issues are difficult to address.
Observations to date support the existence of a nanometer-thick premelted layer in pure materials a few degrees below the bulk melting point and there is more ample evidence for premelting in alloys.
Modeling activities based on PFC \cite{Mellenthinetal2008} and MD simulations \cite{Hoytetal2009,Fensinetal2010} allow characterization of the structural forces underlying GB premelting.

\begin{figure}
\begin{center}
\includegraphics[width=8cm]{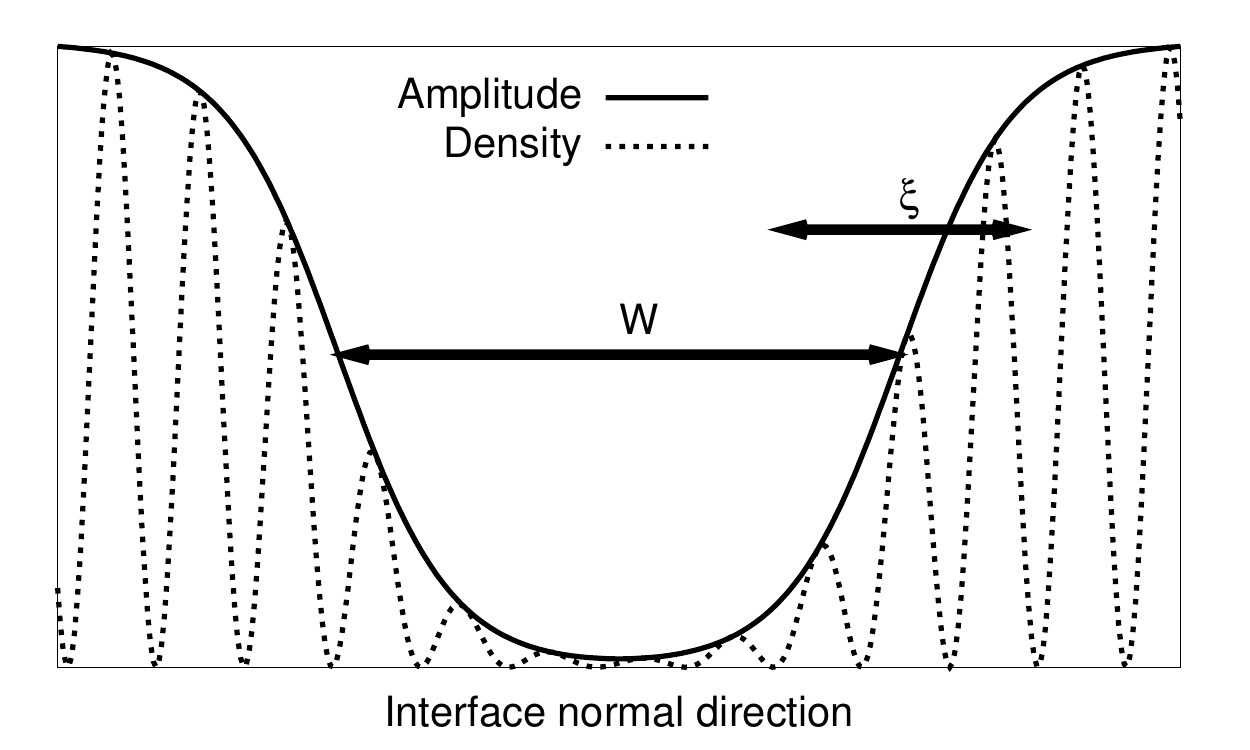}
\caption{Sketch of the atomic density profile at two adjacent solid-melt interfaces and the corresponding amplitude.}
\label{intro::sketch}
\end{center}
\end{figure}

These forces are quantified by the introduction of the ``disjoining potential'' $V(W)$, which is defined via the excess Gibbs free-energy per unit of grain boundary area 
\begin{equation}
G_{\rm exc}(W,T)=\Delta G(T)W+2\gamma_{sl}+V(W),
\label{gbexc}
\end{equation}
where $\Delta G=G_s-G_l$ is the bulk Gibbs free-energy difference between liquid ($G_l(T)$) and solid ($G_s(T)$) and $\gamma_{sl}$ is the solid-liquid interfacial free-energy. 
Based on this definition, $V(W)$ represents the part of this excess due to the overlap of crystal density waves from the two grains on each side of the GB (see Fig.~\ref{intro::sketch}).
The ordering of the solid phases extends also into the melt on the range of the interface thickness $\xi$, thus the crystals start to interact with each other as soon as their separation $W$ is of the order of the interface thickness.
Depending on the alignment of the crystals their structures may match --- which leads to an attractive interaction --- or are locally displaced such that the energy of the system is increased by the overlap, which leads to a repulsive interaction.
Hence the derivative $-dV(W)/dW$ expresses the force between crystal-melt interfaces due to this overlap, which can be either repulsive or attractive depending on whether the sign of $-dV(W)/dW$ is positive or negative, respectively.
So far, there is little analytical knowledge on the short range contributions to these forces, with the exception of phase-field models \cite{Wangetal2008}, which are based on phenomenological models that do not consider atomic structures, dislocation formation and elastic interactions.
The purpose of the present article is therefore to gain analytical insights into the nature of these forces, based on a complex Ginzburg-Landau description.
Additionally, $V(W)$ also contains attractive contributions due to London dispersion forces that are neither accounted for in PFC and MD simulations, nor in the short range forces analyzed here, but play an important role in other systems such as ceramic materials \cite{Clarke1987}.
However, in metallic systems, dispersion forces can be estimated to only contribute an attractive tail to $V(W)$ whose magnitude is less than a mJ/m$^2$ for $W$ on the nanometer scale.
In contrast, MD computations of  $V(W)$ in pure Ni \cite{Hoytetal2009,Fensinetal2010} show that $V(W)$ has a magnitude of tens of mJ/m$^2$ for $W$ in this same range.

A major outcome of this work is that the structural interaction can be calculated analytically 
in the case of zero misorientation between the grains, which only have a translational misfit.
The range of the interaction can still be computed also for the more general case of misoriented crystals.
The results therefore offer new insights into the phenomenon of GB premelting, as they show which quantities and ingredients are essential for the structural interactions.
This paper therefore complements the numerical results in Ref.~\onlinecite{TheNeverendingStory}, which compare PFC and amplitude equations results with MD data.

The structure of the article is as follows:
In Section \ref{ae} the underlying model is summarized, which is then used in Section \ref{sl} to investigate the properties of single solid-melt interfaces.
There, in particular, the decay of the density waves into the melt is analyzed, since this turns out to be the key parameter for the interaction of two solid-melt interfaces.
First, in Section \ref{InterfaceInteraction} the special case of crystals without misorientation but with translational misfit is considered, as here the asymptotics of the interaction can be calculated fully analytically, which is demonstrated for several interface orientations.
In Section \ref{misoriented} grain boundaries are considered;
although a full analytical treatment is not possible here, still the range of the interaction can be predicted.

\section{Amplitude equations}
\label{ae}

From the classical density functional theory of freezing (DFT) a functional can be derived which expresses an emerging solid phase as density fluctuations appearing from the liquid state, whereas the (time-averaged) density is spatially constant in the melt phase \cite{HaymetOxtoby1981, Lairdetal1987, Singh1991, HarwellOxtoby1984, ShenOxtoby1996a, ShenOxtoby1996b, Wuetal06, Wu07, SpaKar09}.
To this end the spatial variations of the density field $\delta \psi(\vec{r})$, are expanded as a sum of density waves
\begin{equation}
\delta\psi(\vec{r})=\sum_{j=1}^N \uj e^{i\vec{k}^{(j)}\cdot\vec{r}},
\label{cexp}
\end{equation}
where each $\vec{k}_i$ is one of the $N$ different reciprocal lattice vectors (RLVs) and $\uj$ are their associated amplitudes.
In the liquid phase, where the time averaged density is spatially constant, the amplitudes vanish, and in an undistorted solid phase they all attain the same constant value, $\uj=u_s$.
The associated free energy deviation from the liquid state is
\begin{eqnarray} 
F &=& \int d\rv \Big( \frac{n_0 k_B T}{2} \sum_{j=1}^N \left[ \frac{\uj \ujs}{S(q_0)}- \frac{C''(q_0)}{2} \left|\square_j \uj\right|^2 \right] \nonumber \\ 
&& + f(\{\uj\}, T) \Big), \label{ae::eq2}
\end{eqnarray}
where the function $f(\{\uj\}, T)$ contains the higher order nonlinear terms in the amplitudes $\uj$ and an explicit dependence on the temperature $T$.
Furthermore, $n_0$ is the density in the liquid state and $C(r)$ is the direct correlation function with Fourier transform
\begin{equation}
C(q) = n_0 \int d\rv\, C(r) \exp(-i\kv\cdot\rv)
\end{equation}
with $r=|\rv|$ and $q=|\kv|$;
it is related to the liquid structure factor by
\begin{equation}
S(q) = \frac{1}{1-C(q)}.
\end{equation}
Here, all quantities are evaluated at the (first) peak of the structure factor $q_0$.
Unlike DFT, where a large number of modes is required to obtain sharp peaks around atomic positions, the simpler free energy allows for a truncation of this sum to a small set of reciprocal lattice vectors.  Various methods have been developed to change the kernel of the free energy in order to stabilize a variety of two and three dimensional periodic and crystal structures.
Here we focus on bcc structures, therefore we restrict the summation to the $N=12$ principal reciprocal vectors
\begin{eqnarray}
&&[110], [101], [011], [1\bar{1}0], [10\bar{1}], [01\bar{1}] \nonumber \\
&&[\bar{1}\bar{1}0], [\bar{1}0\bar{1}], [0\bar{1}\bar{1}], [\bar{1}10], [\bar{1}01], [0\bar{1}1].
\end{eqnarray}
Notice that by the condition of having a real density field $\psi$ the $N$ complex amplitudes are not independent but are complex conjugate (denoted by a star) for antiparallel RLVs.
Therefore, we restrict the description to the first row of these RLVs and use only $N/2$ independent complex fields $\uj$.

The differential operator $\Box_j$ is given by \cite{Gunaratne, Graham, SpaKar09}
\begin{equation}
\Box_j = \khj\cdot \nabla - \frac{i}{2q_0} \nabla^2,
\end{equation}
where the vectors $\khj$ are the normalized principal RLVs.
The second term in the operator preserves the rotational invariance of the equations and is related to the use of the nonlinear strain tensor in elasticity.
Below we refer to this second term as the higher order correction term in the box operator, since it is vanishingly small for rough interfaces.


For most of the present analysis the precise form of the higher order nonlinearities is not important, as they enter the expressions for the interface interaction only as prefactors in terms of matching constants.
However, to complete the model, we use here amplitude equations which are derived via a multiscale expansion from the three-dimensional phase-field-crystal model \cite{SpaKar09},
\begin{equation}
{\cal F} = \int d\vec{r} \left( \frac{\psi}{2} [-\epsilon + (\nabla^2 +1)^2]\psi + \frac{1}{4}\psi^4 \right).
\end{equation}
Here, $\epsilon$ is used as a small parameter in the regime of the phase diagram which describes the coexistence between the bcc and the homogeneous (melt) phase.
The parameter $\epsilon$ is related to the physical parameters via
\begin{equation}
\epsilon = \frac{103}{96} \epsnew 
\end{equation}
with another (small) dimensionless parameter $\epsnew$
\begin{equation}
\epsnew = - \frac{24}{S(q_0) C''(q_0) q_0^2},
\end{equation}
which turns out to be more useful in the context of the amplitude equations;
it characterizes the ratio between the square of the atomic spacing and the solid-liquid interface thickness.

In equilibrium the chemical potential
\begin{equation}
\mu = \frac{\delta {\cal F}}{\delta \psi} = -\epsilon \psi + (\nabla^2 +1)^2\psi + \psi^3
\end{equation}
is spatially constant.
A detailed derivation of the amplitude equations, which describe the evolution of the fields $u^{(j)}$ has been given in Refs.~\onlinecite{Wuetal06, SpaKar09}, and therefore we only give the resulting expressions here.
The evolution equations can be derived from a free energy functional, which reads
\begin{eqnarray}
F &=& F_0 \int d\Rv \Bigg[ \sum_{i=1}^{N/2} |\Box_j \Aj |^2 + \frac{1}{12} \sum_{j=1}^{N/2} \Aj\Ajs \nonumber \\
&& + \frac{1}{90} \Bigg\{ \left( \sum_{j=1}^{N/2} \Aj\Ajs \right)^2  - \frac{1}{2}\sum_{j=1}^{N/2} |\Aj|^4 \nonumber \\
&& + 2A_{110}^* A_{1\bar{1}0}^* A_{101} A_{10\bar{1}} + 2A_{110} A_{1\bar{1}0} A_{101}^* A_{10\bar{1}}^* \nonumber \\
&& + 2A_{1\bar{1}0} A_{011} A_{01\bar{1}} A_{110}^* + 2A_{1\bar{1}0}^* A_{011}^* A_{01\bar{1}}^* A_{110} \nonumber \\
&& + 2A_{01\bar{1}} A_{10\bar{1}}^* A_{101} A_{011}^* + 2A_{01\bar{1}}^* A_{10\bar{1}} A_{101}^* A_{011} \Bigg\} \nonumber \\
&& -\frac{1}{8} \big( A_{011}^* A_{101} A_{1\bar{1}0}^* + A_{011} A_{101}^* A_{1\bar{1}0} + A_{011}^* A_{110} A_{10\bar{1}}^* \nonumber \\
&& + A_{011} A_{110}^* A_{10\bar{1}} + A_{01\bar{1}}^* A_{110} A_{101}^* + A_{01\bar{1}} A_{110}^* A_{101} \nonumber \\
&& + A_{01\bar{1}}^* A_{10\bar{1}} A_{1\bar{1}0}^* + A_{01\bar{1}} A_{10\bar{1}}^* A_{1\bar{1}0} \big) \Bigg] + F_T. \label{fenAE}
\end{eqnarray}
Here, we have introduced rescaled amplitudes
\begin{equation}
\Aj = \uj/u_s.
\end{equation}
Also, we have introduced a dimensionless length scale
\begin{equation}
\Rv =\epsnew^{1/2}q_0 \rv.
\end{equation}
In these rescaled coordinates the box operator becomes
\begin{equation}
\Box_j = \khj\cdot \nabla - \frac{i\epsnew^{1/2}}{2} \nabla^2,
\end{equation}
and the prefactor $F_0$ is
\begin{equation}
F_0 = - \frac{n_0 k_B T}{2} C''(q_0) q_0^{-1} u_s^2 \epsnew^{-1/2}.
\end{equation}
Finally, the thermal tilt $F_T$ is added phenomenologically,
\begin{equation} \label{tilt::eq1}
F_T = \epsnew^{-3/2} q_0^{-3} \int d\Rv L \frac{T-T_M}{T_M} \phi(\{\Aj\})
\end{equation}
to favor either the solid or liquid state.
Here, $T_M$ is the melting temperature, $L$ the latent heat and $\phi$ is an ``order parameter'' which discriminates between solid and liquid,
\begin{equation} \label{tilt::eq2}
\phi(\{\Aj\}) = \frac{2}{N} \sum_{j=1}^{N/2} h(|\Aj|^2)
\end{equation}
with
\begin{equation} \label{tilt::eq2a}
h(\phi) = \phi^2(3-2\phi).
\end{equation}


Alternatively, the tilt can be chosen such that it reproduces the PFC results, and then the coupling function is chosen to be
\begin{equation} \label{tilt::eq3}
\phi(\{\Aj\}) = \frac{2}{N} \sum_{j=1}^{N/2}     \sqrt{\Aj \Ajs}.
\end{equation}
We note that this expression is invariant under elastic deformations and lattice rotations, which affect the complex phases of the amplitudes.
In the original DFT formulation the coupling therefore reads
\begin{equation} \label{tilt::eq4}
F_T = L \frac{T-T_M}{T_M} \int d\rv  \sum_{j=1}^{N/2} \frac{2\sqrt{\uj \ujs}}{N u_s}.
\end{equation}

We note that the two above coupling functions (\ref{tilt::eq2}) and (\ref{tilt::eq3}) are substantially different:
The first is quartic in the amplitudes variations in the solid and liquid state, and therefore the minima of the functional remain at $\Aj=0$ and $\Aj=1$ for $T\neq T_M$.
This is not the case for the second coupling, which is linear in the amplitudes;
therefore here the bulk values depend on the temperature.
We will discuss the implications of these two different couplings in Appendix \ref{lincoupling}.

Thermodynamic equilibrium corresponds to a stationary state of the free energy functional, and we use relaxation dynamics according to
\begin{equation} \label{relaxscheme}
\frac{\partial \Aj}{\partial t} = - K_j \frac{\delta F}{\delta \Ajs}.
\end{equation}
Since we focus here exclusively on static situations, the choice of the kinetic coefficients $K_j$ is arbitrary.

This description predicts the correct anisotropy of surfaces energies \cite{Wuetal06} and elastic properties and contains naturally the linear theory of elasticity \cite{SpaKar09}.
The form of these nonlinearities depends slightly on the underlying model:
Above it is given for a PFC model, and there are some differences if these terms are derived from DFT using an equal weight ansatz.
However, the differences are small and lead e.g.~only to tiny changes of the anisotropy of the surface energy, as had been investigated in Ref.~\onlinecite{Wu07}.
Also, we point out that --- as will be shown in the following sections --- the higher order nonlinearities do not contribute to the short-range interaction tail for shifted crystals.

Finally, we note that this amplitude equations model for crystals is conceptually close to theories of superconductivity and pattern formation in hydrodynamics \cite{Cross}.

\section{Solid-liquid interfaces}
\label{sl}

Properties of solid-melt interfaces, in particular interfacial energies and their anisotropy, were discussed in detail in Ref.~\onlinecite{Wuetal06,Wu07}.
Here we concentrate on specific properties that are relevant for the understanding of interface interactions in the next section.

In the melt region sufficiently far away from the interface, the amplitudes of the density waves have decayed and can be well described by the linearized equilibrium conditions or, equivalently, the free-energy density with the local terms up to second order in the amplitudes.
It is worthwhile to mention that the free energy functionals, as derived from PFC and DFT, agree up to this order, and therefore the following results are generic.

Let the interface normal $\vec{n}$ be the $z$ direction of a straight solid-liquid interface, and all density wave amplitudes depend then only on this coordinate.
In the liquid, where the solid ordering has decayed almost completely, the equilibrium conditions decouple and are given by
\begin{equation} \label{single::eq1}
\frac{1}{S(q_0)} \uj = -\frac{1}{2} C''(q_0) (\khj\cdot\vec{n})^2 {\dduj},
\end{equation}
where we ignore for the moment the higher order corrections of the box operator.
For reasons that will become more obvious later, we denote here derivatives with respect to $z$ by a dot.
Although we consider a three-dimensional situation, the amplitudes depend only on $z$.
We focus here on stationary states, therefore time-derivatives do not appear.
Obviously, (stationary) coexistence between solid and liquid bulk phases with a planar interface is only possible for $T=T_M$.

The general solution of this linearized equilibrium condition is a superposition of two exponentials,
\begin{equation} \label{single::eq2}
\uj = c_{j, in} \exp (-\lambda_j z) + c_{j, out} \exp(\lambda_j z)
\end{equation}
with the inverse decay length
\begin{equation} \label{single::eq3}
\lambda_j = \left( \frac{-2}{S(q_0) C''(q_0) (\khj\cdot\hat{n})^2} \right)^{1/2}.
\end{equation}
We also define the characteristic scale $\lambda_0\sim 1/\xi$
\begin{equation}  \label{single::eq4}
\lambda_0 = \left( \frac{-2}{S(q_0) C''(q_0)}\right)^{1/2}.
\end{equation}
Since we consider only a single interface, with the solid phase being located at $z\to -\infty$, the growing exponential cannot be present ($c_{j, out}=0$). 
We note that a shift of the interface by $\Delta z$ leads to a change of the remaining exponential prefactor $c_{j, in}$ by a factor $\exp(-\lambda_j\Delta z)$.

Since the problem is one-dimensional, it is straightforward and fast to solve the full set of amplitude equations (not only in the linearized region) using a real space implementation via the relaxation scheme (\ref{relaxscheme}) at $T=T_M$, until an equilibrium solid-liquid interface is established.
The grid spacing is chosen to be much smaller than the interface thickness, $\lambda_j dz \ll 1$, to obtain results which do not depend on the discretization.
Corresponding to the analytical investigation we do not take into account the higher order term in the box operator, thus the equilibrium profile is described by a second order ordinary differential equation.
From the equilibrium profile the solid-liquid interfacial energy $\gamma_{sl}$ is also computed.
To obtain a numerical value for the prefactor $c_{j, in}$ we have to match it to the full solution of the nonlinear problem $\delta F/\delta u^{(i)}=0$.
Therefore, we set the origin $z=0$ exactly at the location of the interface.
Since the interface is smooth, the position of it requires a precise definition.
The choice of this measure is not critical, since another definition only leads to slightly different values for the exponential prefactors, and later on in the following sections, to a horizontal shift of the disjoining potential.
We use an integral measure for the amount of liquid per unit area of the interface
\begin{equation}  \label{single::eq5}
W(L_z) = \frac{1}{N} \int_0^{L_z} dz\, \sum_{j=1}^N [1-h(|\uj|^2/u_s^2)],
\end{equation}
where we use the ``coupling function'' (\ref{tilt::eq2a}), $h(x)=x^2(3-2x)$ to interpolate between solid and liquid.
Notice that in the liquid the amplitudes have decayed to zero, whereas in the solid all of them have the value $|\uj|=u_s$.

\begin{figure}
\begin{center}
\includegraphics[width=8cm]{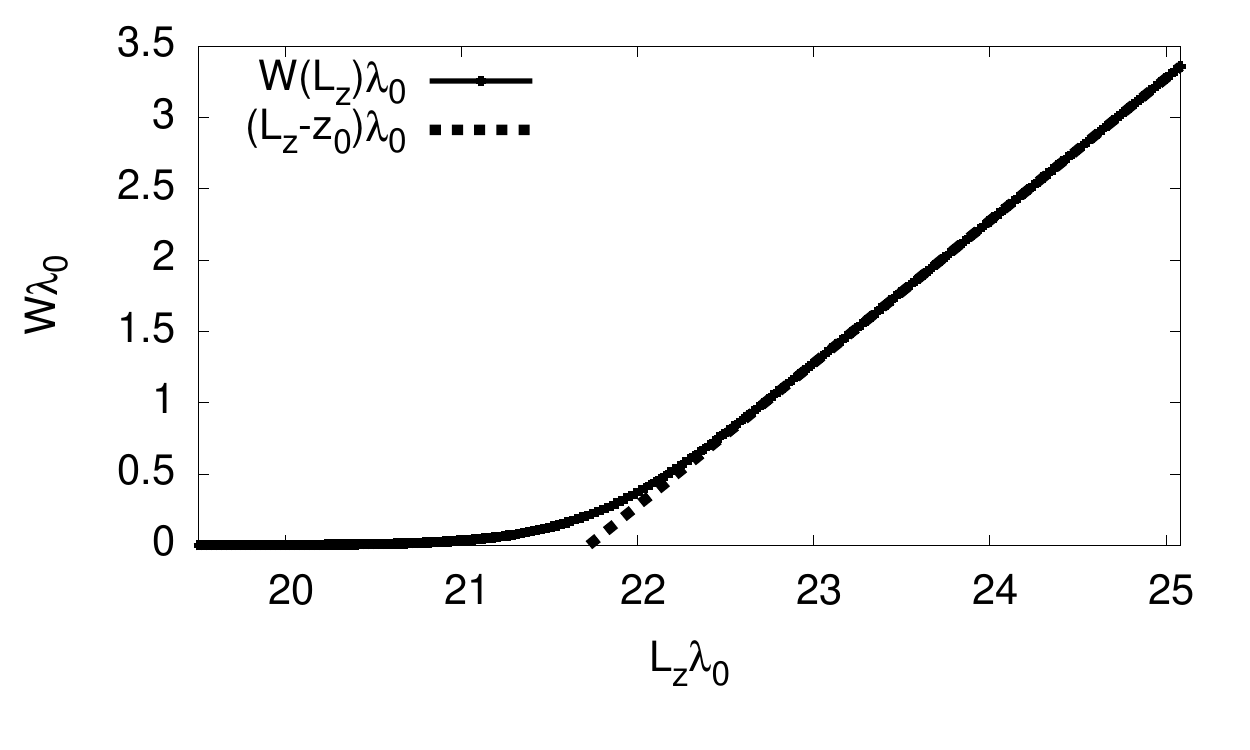}
\caption{Determination of the interface position. The solid phase is located in the right, the liquid in the left half of the system. The amount of melt is defined through the measure (\ref{single::eq5}). For sufficiently large systems, this expression becomes asymptotically equal to $L_z-z_0$, where $L_z$ is the system length and $z_0$ the interface position. Results are shown here for a $(100)$ interface at $T=T_M$.
In the present case, the interface is located at $z_0\lambda_0=21.7$.
}
\label{single::fig1}
\end{center}
\end{figure}

For a single solid-liquid interface, the amount of liquid depends of course on the system size, i.e.~the length of the integration interval $L_z$.
For $L_z\gg z_0$ ($z_0$ is the interface position), $W$ becomes a linear function of $L_z$, $W\simeq L_z-z_0$.
We can extrapolate this linear function to the value $0$, which then defines the location of the interface, and this is shown in Fig.~\ref{single::fig1}.

\begin{figure}
\begin{center}
\includegraphics[width=8cm]{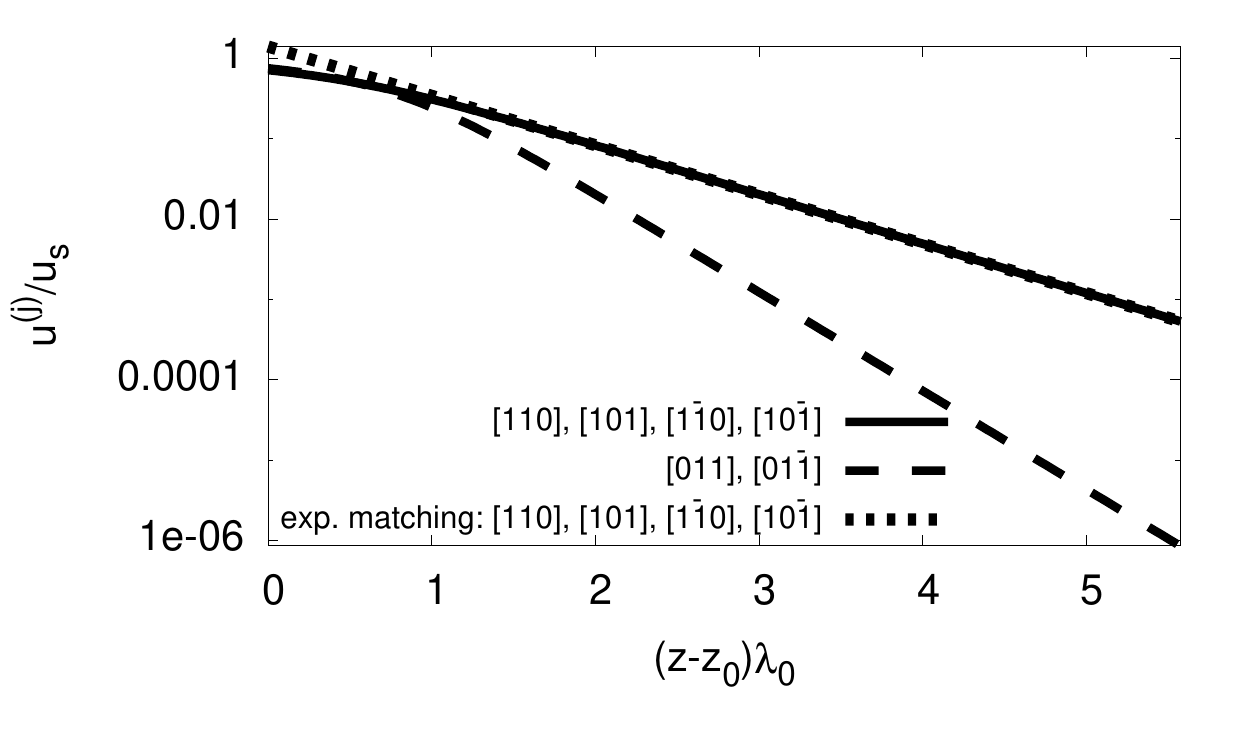}
\caption{Matching of the exponential decay of the amplitudes. For a (100) surface, the amplitudes group into two classes.
The matching constants are determined such that the curves for the slowest decaying density wave amplitudes (solid curve) agree with the exponential solution of the linearized equations (dotted line).
}
\label{single::fig3}
\end{center}
\end{figure}

In the next step, we plot the amplitudes as a function of the distance from the interface, $z-z_0$.
For $z\gg z_0$ they decay exponentially on the scale $1/\lambda_j$, and we can determine the exponential prefactors as shown in Fig.~\ref{single::fig3}.
We note that for a single interface all amplitudes can be chosen to be purely real (as long as the correction from the box operator is suppressed).
For a (100) surface, only the density waves $[110]$, $[1\bar{1}0]$, $[101]$, $[10\bar{1}]$ (+ complex conjugates) follow the exponential decay given above, since for the others the interface normal is perpendicular to the principal reciprocal lattice vectors, $\kh\cdot\hat{n}=0$.
This means that they decay faster, as they are ``slaved'' by the other fields (see Fig.~\ref{single::fig3}).
As explained in Ref.~\onlinecite{Wuetal06} the density waves can therefore be grouped into two classes.
The matching constant for the slowly decaying fields is then determined numerically as $c_{110}^{(100)}=c_{1\bar{1}0}^{(100)}=c_{101}^{(100)}=c_{10\bar{1}}^{(100)}=0.165$ for the definition of the interface position given above (the subscript denotes the density wave, the superscript the interface normal).

\begin{figure}
\begin{center}
\includegraphics[width=8cm]{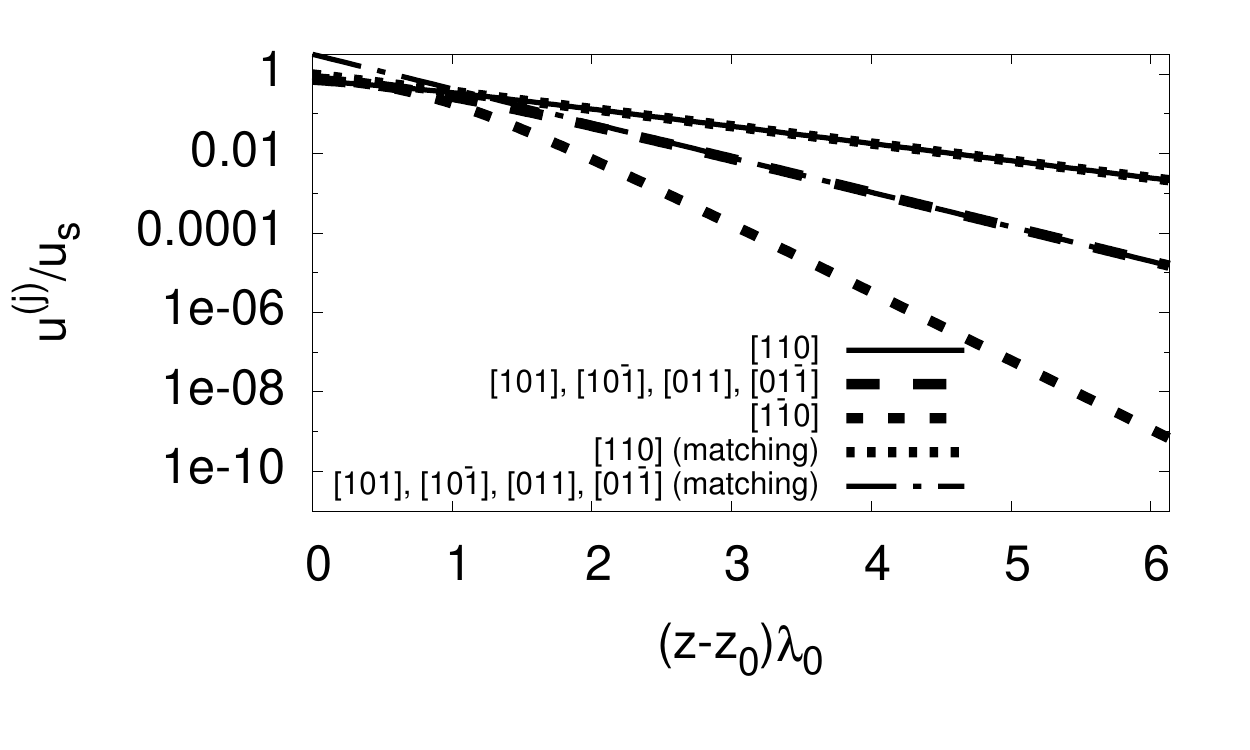}
\caption{Matching of the exponential decay of the amplitudes. For a (110) surface, the amplitudes group into three classes: $[110]$ has the longest range, the second group is $[101], [10\bar{1}], [011], [01\bar{1}]$. The shortest range density wave is related to the wave vector $[1\bar{1}0]$.
}
\label{single::fig4}
\end{center}
\end{figure}

The same procedure can be repeated for any other interface, and in general all matching constant are different (they are only pairwise equal for complex conjugate fields).
The corresponding plot for a (110) surface is shown in Fig.~\ref{single::fig4}.
The obtained matching constants are $c_{110}^{(110)}=0.116$ and $c_{101}^{(110)}=c_{10\bar{1}}^{(110)}=c_{011}^{(110)}=c_{01\bar{1}}^{(110)}=0.372$;
the remaining field $u_{1\bar{1}0}$ is slaved by the others and decays faster.

\begin{figure}
\begin{center}
\includegraphics[width=8cm]{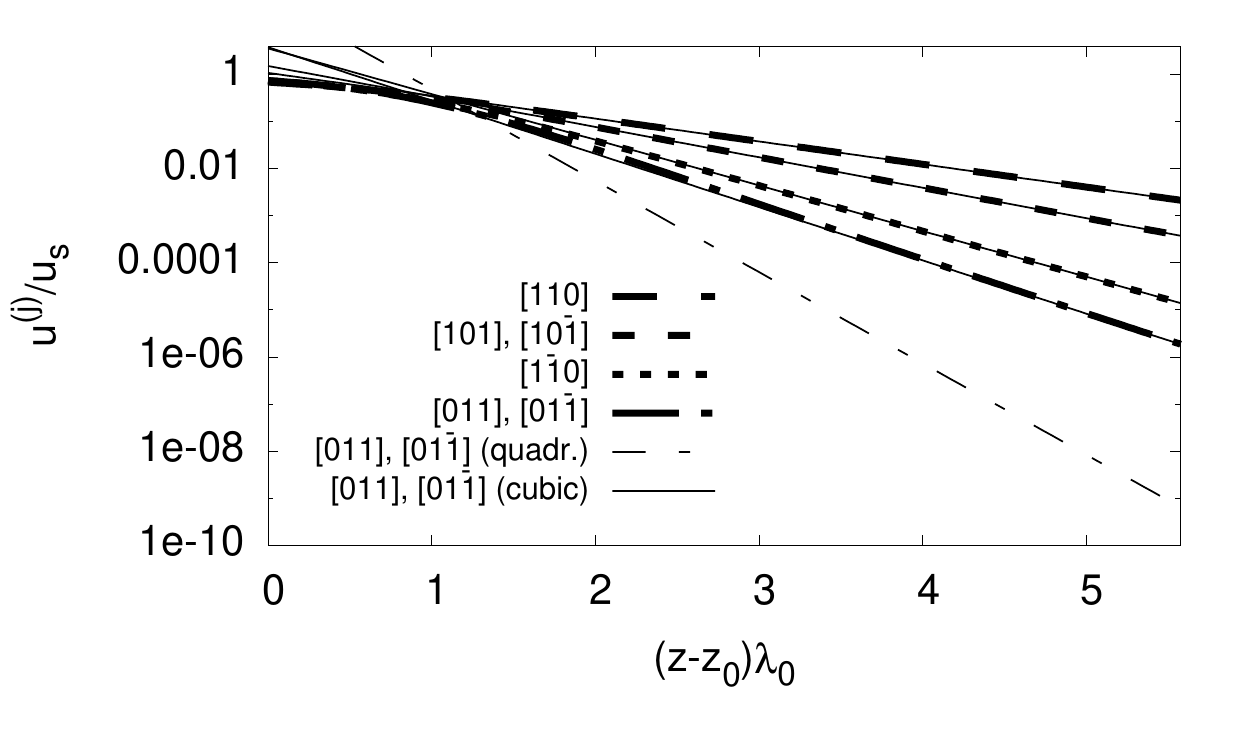}
\caption{Matching of the exponential decay of the amplitudes for a (310) solid-liquid interface.
The range of the fastest decaying field $u_{011}$ is not determined by quadratic but cubic interactions.
}
\label{single::fig5}
\end{center}
\end{figure}

Finally, for a (310) interface (as a representative for an arbitrary interface normal direction), the numerical matching gives:
$c_{110}^{(310)}=0.128$, $c_{101}^{(310)}=c_{10\bar{1}}^{(310)}=0.18$, $c_{1\bar{1}0}^{(310)}=0.425$.
We note that the fields $u_{011}$ and $u_{01\bar{1}}$, which are expected to have a decay according to Eq.~(\ref{single::eq3}), turn out to behave differently;
they decay more slowly than anticipated, because e.g.~for $u_{011}$ a forcing term of the structure $\sim u_{110}u_{10\bar{1}}^*$ from the cubic terms in the free energy functional leads to a longer range of the density wave than the anticipated quadratic term;
in fact, for the given inclination $\lambda_{011}^{(310)} > \lambda_{110}^{(310)}+\lambda_{10\bar{1}}^{(310)}$, with the right hand side being the inverse decay length of the slaved field.
In general, it means that also density wave amplitudes with $\kh\cdot\hat{n}\neq 0$ can be slaved by other terms, and their decay is then determined by the higher order nonlinearities.
The range of these slaved fields is very short, and therefore they do not contribute significantly to the interaction potentials derived below.
The decay of all amplitudes, together with the exponential fits, is shown in Fig.~\ref{single::fig5}.

Let us briefly discuss the relevance of the box operator corrections to the preceding results.
So far, a single straight interface is in principle described by a real density wave amplitude, or at least by a constant complex phase.
The box operator explicitly introduces an imaginary factor, and therefore the amplitudes pick up a small and slow oscillatory contribution.
Also, the results without the higher order corrections (as given on the slow scale), do not yet depend on the value of $\epsnew$, which is only re-introduced when the density profile is reconstructed from the amplitudes.
With the correction terms, $\epsnew$ appears also explicitly in the amplitude equations.
The linearized equations in the liquid region become
\begin{equation} \label{singleint::eq1a}
\frac{1}{S(q_0)} \uj + \frac{1}{2} C''(q_0) \Box_j^2 \uj = 0,
\end{equation}
and in one dimension 
\begin{equation}
\Box_j \uj \rightarrow (\khj\cdot\hat{n}) \duj - \frac{i}{2|\kvj|} \dduj.
\end{equation}
The general solution of the linearized problem then becomes
\begin{eqnarray}
\uj &=& \cjam \exp(\lambda_{j, a}^- z) + \cjbm\exp(\lambda_{j, b}^- z) \nonumber \\
&+&  \cjap \exp(\lambda_{j, a}^+ z) + \cjbp\exp(\lambda_{j, b}^+ z) \label{singleint::eq1},
\end{eqnarray}
with four independent solutions, since the equation is now of fourth order.
The new decay scales are given by
\begin{eqnarray}
\lambda_{j, a}^+ &=& -i(\khj\cdot\hat{n}) |\kvj| \nonumber \\
&& + \left( -(\khj\cdot\hat{n})^2 |\kvj|^2  + i \frac{\sqrt{8} |\kvj|}{\sqrt{-SC''}} \right)^{1/2}, \label{boxrange1} \\
\lambda_{j, b}^+ &=& -i(\khj\cdot\hat{n}) |\kvj| \nonumber \\
&&+ \left( -(\khj\cdot\hat{n})^2 |\kvj|^2  - i \frac{\sqrt{8} |\kvj|}{\sqrt{-SC''}} \right)^{1/2}, \label{boxrange2}\\
\lambda_{j, a}^- &=& -i(\khj\cdot\hat{n}) |\kvj| \nonumber \\
&& - \left( -(\khj\cdot\hat{n})^2 |\kvj|^2  - i \frac{\sqrt{8} |\kvj|}{\sqrt{-SC''}} \right)^{1/2}, \label{boxrange3}\\
\lambda_{j, b}^- &=& -i(\khj\cdot\hat{n}) |\kvj| \nonumber \\
&& - \left( -(\khj\cdot\hat{n})^2 |\kvj|^2  + i \frac{\sqrt{8} |\kvj|}{\sqrt{-SC''}} \right)^{1/2}, \label{boxrange4}
\end{eqnarray}
with the abbreviations $C''=C''(q_0)$ and $S=S(q_0)$.
Hence we have for the real parts the relation $\Re(\lambda_{j, a}^+) = \Re(\lambda_{j, b}^+) = -\Re(\lambda_{j, a}^-) = -\Re(\lambda_{j, b}^-) > 0$, which means that the solutions with superscript $+$ are growing solutions and the ones with $-$ are decaying solutions, all with the same range.
Notice that they all also have oscillatory contributions, i.e.~a non-vanishing imaginary part, $\Im(\lambda) \neq 0$.
We also have obviously $\lambda_{j, a}^+ = -\lambda_{j, a}^{-*}$ and $\lambda_{j, b}^+ = -\lambda_{j, b}^{-*}$.
These relations are important for the proper matching of incoming and outgoing waves in the interface region between the two grains.
They imply $\Im(\lambda_{j, a}^+) = \Im(\lambda_{j, a}^-)$ and $\Im(\lambda_{j, b}^+) = \Im(\lambda_{j, b}^-)$, therefore the oscillation frequency is equal for corresponding decaying and growing solutions.
Also, the growth rates $\lambda_{j, a}^\pm$ are only weakly imaginary in contrast to $\lambda_{j, b}^\pm$.
It is therefore not surprising that we find numerically that amplitudes of the strongly oscillatory solutions are very small, $|c_{j, b}^\pm| \ll |c_{j, a}^\pm|$, since an interface should mainly be a decay and not an oscillation of the density waves -- the latter corresponds to a local change of the lattice spacing;
the oscillatory modes can therefore safely be neglected.
For $\delta$ iron we have $\epsnew=0.0860$ (i.e.~$\epsilon = 0.0923$), and for this value $\lambda$ is only very slightly changed, and the amplitudes almost undistinguishable.
Notice, however, that $\lambda$ formally becomes complex and that the decay rates $\Re(\lambda)$ for the incoming and outgoing waves are slightly different, but for present small values of $\epsnew$ this difference can be neglected.

\section{Interface interaction}
\label{InterfaceInteraction}

\subsection{General framework}

\begin{figure}
\begin{center}
\includegraphics[trim=1cm 8cm 8cm 0cm, clip=true, width=8cm]{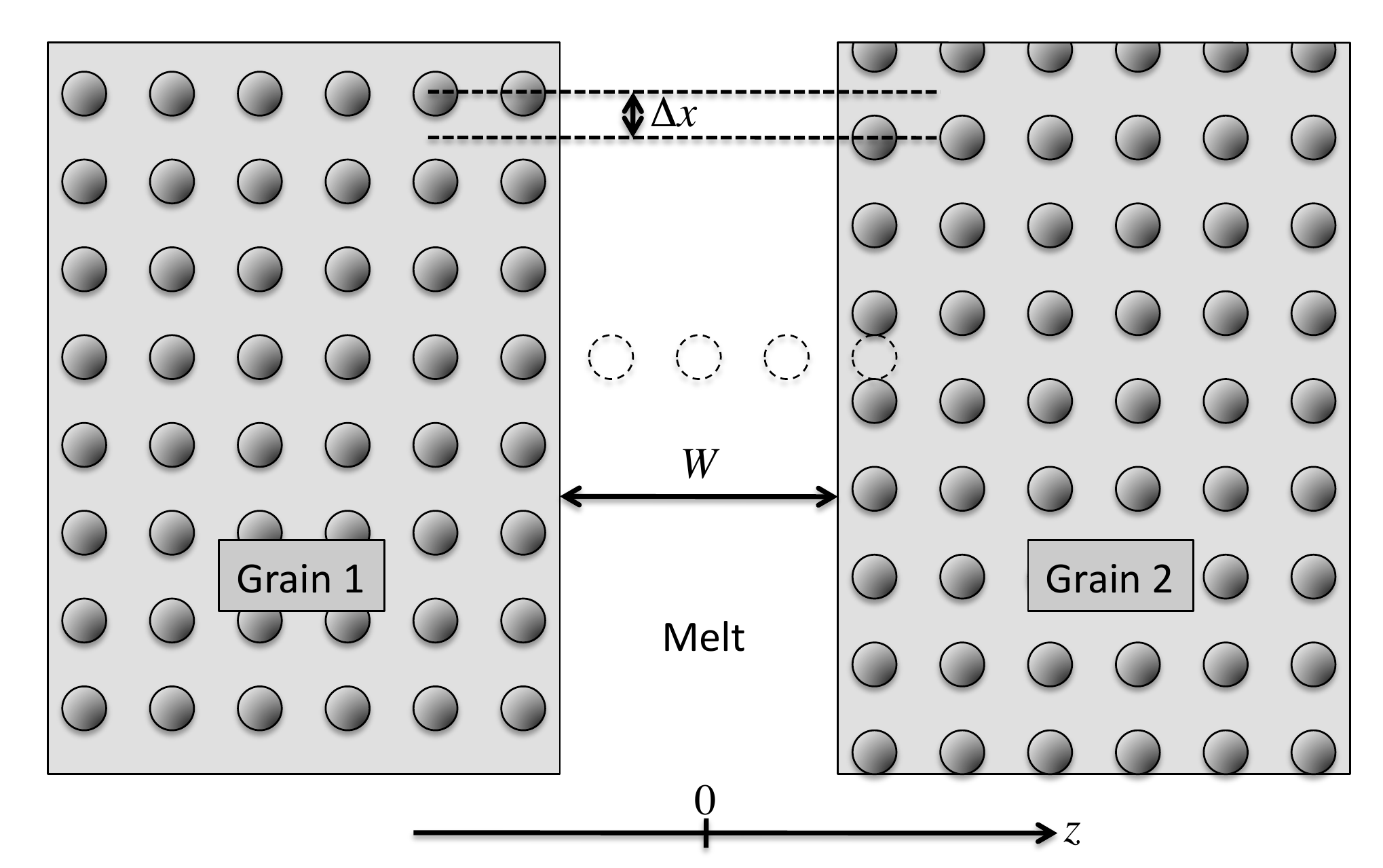}
\caption{Sketch of the geometry for shifted crystals. The displacement in the out-of-plane direction, $\Delta y$, is not shown. We assume that in the normal direction the crystals are not shifted, as illustrated by the dotted circles.}
\label{shift::fig1}
\end{center}
\end{figure}

The simplest case of interacting solids is that of two lattices of the same material and with the same structure that are perfectly aligned up to a translation in the contact plane, i.e.~without misorientation between them (see Fig.~\ref{shift::fig1}).
If the crystals are fully aligned, which means that the atomic planes match, the interaction between the solid-melt interfaces is attractive, because exactly at the melting point ($T=T_M$) merging of the crystals removes two solid-melt interfaces, which reduces the total energy.
On the other hand, the situation is more complicated if the crystals are shifted against each other, which implies elastic deformations of the lattices close to the grain interface.
As we will show, a sufficiently large mismatch can lead to repulsive interactions.

It is quite remarkable that the asymptotic of this structural interaction between the crystals can be calculated fully analytically from the free-energy expression.
The procedure is as follows:
We assume two parallel crystal surfaces (see Fig.~\ref{shift::fig1}), which are separated by a melt layer of thickness $W$.
For large $W$, the density wave amplitudes are almost decayed in the center of the melt, and it is therefore sufficient to consider only the free energy terms up to quadratic order in the amplitudes.
The related equilibrium equations are therefore linear and can be solved easily, and the corresponding (approximative) solution has to be matched to the exact solution of the full problem of a localized interface at $z\pm W/2$.
By the means of this matching, we get an analytical expression for the disjoining potential.

We start the analysis with the derivation of a conservation law.
As before, we first ignore the higher order correction that stems from the box operator.
The full free-energy expressions (\ref{ae::eq2}) and (\ref{fenAE}) have the structure 
\begin{equation} \label{shift::eq1}
F = \int (f_p + f_k) dz,
\end{equation}
where $f_p$ depends only on local terms (no gradients of the amplitudes), whereas $f_k$ contains only first order derivative terms.
Notice that due to the parallel structure, all amplitudes depend only on the coordinate $z$ perpendicular to the interfaces.
Equilibrium demands
\begin{equation} \label{shift::eq2}
\frac{\delta F}{\delta \uj} = 0
\end{equation}
for all fields $\uj(z)$.

For a solid-melt-solid layer system, the free energy is in the spirit of equation (\ref{gbexc})
\begin{equation} \label{fgbexc}
F = - W \Delta f + V(W) + 2\gamma_{sl}
\end{equation}
in the present case of the underlying NVT ensemble.
The bulk free energy density difference $\Delta f=L(T-T_M)/T_M$ for a temperature deviation from the melting temperature $T_M$ corresponds to $-\Delta G$ introduced in Eq.~(\ref{gbexc}) and will be discussed in detail below.

To emphasize the analogy to a problem in classical mechanics, we use a dot for the spatial derivative in $z$ direction.
The ``Hamiltonian'',
\begin{equation} \label{shift::eq2a}
H=f_k-f_p,
\end{equation}
is then a ``constant of motion'', i.e.~it does not depend on the $z$ coordinate,
\begin{equation} \label{shift::eq2b}
\dot{H} = 0.
\end{equation}
For interfaces that are far apart, the amplitudes have almost decayed to zero in the melt region, and all contributions which are higher than qudratic in the amplitudes give only negligible corrections.
Therefore, we get
\begin{eqnarray}
H &=& -n_0 k_B T \sum_j^{N/2} \Big( \frac{1}{S(q_0)} \uj \ujs \nonumber \\
&& + \frac{1}{2} C''(q_0) (\khj\cdot\hat{n})^2 \duj {\dujs} \Big). \label{shift::eq3}
\end{eqnarray}
The corresponding linearized ``equations of motion'' which describe the small amplitudes in the liquid region are therefore again given by Eqs.~(\ref{single::eq1})-(\ref{single::eq3}), with the only difference that we have here two interfaces, and therefore both exponentials are present.
From this solution we can calculate the Hamiltonian in a quadratic approximation,
\begin{equation} \label{shift::eq7}
H = -2n_0 k_B T  \frac{1}{S(q_0)} \sum_j^{N/2} (c_{j,in}c_{j, out}^* + c_{j,in}^*c_{j, out}).
\end{equation}
We can choose the origin $z=0$ in the center between the two interfaces, and then the exponential prefactors have the same absolute value but can differ by their phase, $c_{j, out} = c_{j, in}\exp(i\phi_j)$.
Furthermore, from the general solution (\ref{single::eq2}) it is obvious that a translation of the interface position in the normal direction increases or decreases the prefactors $c_j$ by an exponential factor $\exp(\lambda_j \Delta z)$, where $\Delta z$ is the shift distance.
Therefore, we get $c_{j, in} = c_{j, 0}\exp(-\lambda_j W/2)$;
the matching constants $c_{j, 0}$ were determined already in the previous section.
Hence,
\begin{equation} \label{shift::eq8}
H = -4 n_0 k_B T \frac{1}{S(q_0)} \sum_j^{N/2} |c_{j, 0}|^2 \exp(-\lambda_j W) \cos\phi_j.
\end{equation}
In general, it is necessary to introduce a tilt term to favor either the liquid or the solid state, because otherwise a repulsive or attractive interaction between the interfaces would forbid the existence of a stationary solution (stable or unstable) with a specific melt layer thickness $W$.
We therefore have to raise or lower the free energy density of the solid phase relative to the liquid by $\Delta f=L(T-T_M)/T_M$.
In particular, overheating above the melting point corresponds to $\Delta f>0$.
Notice that for the following calculation of the asymptotic interface interaction the precise form of the coupling in $f_T$ is not important, and only the tilt $\Delta f$ enters into the result, provided that the bulk states $\uj=0$ and $\uj=u_s$ are temperature independent.
This is the case e.g. for the coupling function (\ref{tilt::eq2a}), or --- more generally --- if the coupling function does not have a linear term in the amplitude variation $\delta \uj = \uj -u_s$ in the solid and $\delta \uj = \uj$ in the melt phase.
The case that a linear term exists will be discussed in more detail below in Appendix \ref{lincoupling}.

In the solid, the amplitudes are (up to a phase factor) all equal to $u_s$, the gradients vanish, and therefore the Hamiltonian becomes
\begin{equation} \label{shift::eq9}
H=-\Delta f.
\end{equation}
Comparison of this exact value, calculated from the solid phase, and the asymptotic value for large interface separations (\ref{shift::eq8}), calculated from the liquid phase, using the conservation law (\ref{shift::eq2b}) we obtain an implicit relation for the (asymptotic) width of the liquid layer $W$ as function of the deviation from the melting point, $\Delta f$.
Asymptotically, only the slowest decaying density waves with the smallest inverse decay length $\lambda_{min}$ contribute to the Hamiltonian, and we get
\[
-\frac{4 n_0 k_B T}{S(q_0)} |c_{min, 0}|^2 \exp(-\lambda_{min} W) \cos\phi_{min} \simeq -L\frac{T-T_M}{T_M},
\]
thus
\begin{equation} \label{shift::eq10}
W \simeq -\frac{1}{\lambda_{min}} \ln \left( \frac{S(q_0)}{4n_0k_B T |c_{min, 0}|^2 \cos\phi_{min}} \frac{L(T-T_M)}{T_M} \right).
\end{equation}
This expression diverges logarithmically at the melting point, where $W=\infty$ is the equilibrium solution.
If $\cos\phi_{min}$ is positive, we find an asymptotic solution only for $T>T_M$.
The interfaces {\em attract} each other, and this has to be compensated by overheating, i.e.~favoring the liquid phase.
On the other hand, for $\cos\phi_{min}<0$, we have {\em repulsive} solutions asymptotically only below the melting point. 

At shorter distances the other density waves also contribute, and we therefore have to sum over all of them, which leads to an implicit relation for the melt layer thickness as function of temperature, 
\begin{equation}
4 n_0 k_B T \frac{1}{S(q_0)} \sum_j^{N/2} |c_{j, 0}|^2 \exp(-\lambda_j W) \cos\phi_j \simeq L \frac{T-T_M}{T_M},
\end{equation}
which follows directly from Eqs.~(\ref{shift::eq8}) and (\ref{shift::eq9}).
This expression is valid as long as the overlap of the density waves is still small, such that the nonlinear energy contributions can be neglected.

We can interpret the free energy shift $\Delta f$ as the chemical force that balances the interface interaction.
In fact, for a single interface it is the driving force for melting or solidification.
From the equilibrium condition $F'(W)=0$ we get by means of Eq.~(\ref{fgbexc})
\begin{equation} \label{shift::eq11}
H = -V'(W) = -\Delta f,
\end{equation}
where $-V'(W)$ is the disjoining force, which is derived from the disjoining potential $V(W)$.
Integrating therefore gives
\begin{eqnarray} \label{shift::eq12}
V(W) &\simeq& -2 n_0 k_B T \sqrt{\frac{-2C''(q_0)}{S(q_0)}} \times \nonumber \\
&\times& \sum_j^{N/2} |\khj\cdot\hat{n}| |c_{j, 0}|^2 \cos\phi_j \exp(-\lambda_j W),
\end{eqnarray}
where we normalized the potential such that it decays to zero for infinitely far separated interfaces, in agreement with Eq.~(\ref{fgbexc}).
Eq.~(\ref{shift::eq12}) is the central result of this article.
Notice that the above expression of the disjoining potential is valid asymptotically for $W\to\infty$.
In this limit $W\lambda_j\gg 1$, which means that the interface thickness is small in comparison to the grain separation $W$.
Then the interfaces are sharp, and the melt layer thickness is (uniquely) well-defined.
For shorter distances, we use the same measure for $W$ as defined above in Eq.~(\ref{single::eq5}), taking into account that for the shifted crystals the interfaces remain planar (the amplitudes depend only on $z$).

The solution of the linear equations is only valid for ``non-slaved'' fields, in particular those with $\khj\cdot\hat{n}\neq 0$, and therefore these fields contribute differently to the disjoining potential by higher order nonlinearities.
However, in the above expression (\ref{shift::eq12}), the fast decaying and therefore negligible fields do not contribute due to orthogonality, $\khj\cdot\hat{n}=0$.

We can choose the origin of the coordinate system such that the amplitudes of one crystal are purely real.
We assume that the other crystal is translated against it in the plane of the grain boundary, so (for a three-dimensional system) we have two translational degrees of freedom.
The translation vector, $\Delta \rv$ then obeys $\Delta \rv\cdot \nh=0$, so with $\nh=\zh$ we get $\Delta \rv = \Delta x \xh + \Delta y \yh$.
The original non-shifted crystal is described by the expression 

\begin{equation}
\delta n(\vec{r})=n_0\sum_j u^{(j)}(\vec{r})\exp(i\vec{k}^{(j)}\cdot\vec{r}),
\end{equation} 
and a translation is therefore described by
\begin{eqnarray*}
\delta n(\rv) &=& n_0 \sum_j \uj (\rv) \exp[i\kvj\cdot (\rv+\Delta \rv)]  \\
&=& n_0 \sum_j \uj (\rv) \exp[i\kvj\cdot \Delta \rv] \exp(i\kvj\cdot \rv).
\end{eqnarray*}
The complex shift factors are therefore given by
\begin{equation}
\phi_j = \kvj\cdot \Delta \rv.
\end{equation}
We define the lateral dependence of the disjoining potential for the fields with equal decay length $\lambda$, or equivalently the same value $\kh\cdot\nh$, in agreement with Eq.~(\ref{shift::eq12}) 
\begin{equation}
f_{\kh\cdot\nh}(\Delta x, \Delta y) = - \sum_{j,\khj\cdot\nh=\kh\cdot\nh}^{N/2} \cos\phi_j,
\end{equation}
where we sum over all amplitudes $j$ with equal decay length.

All density waves with the same decay length, i.e.~equal value of $\kh\cdot\nh$ and $\lambda_j=\lambda$, have the same exponential decay, and we can define
\begin{equation}
V_{\kh\cdot\nh}(W) = 2n_0 k_B T  \sqrt{\frac{-2C''(q_0)}{S(q_0)}} | \kh\cdot\nh | \exp(-\lambda W).
\end{equation}
The disjoining potential therefore becomes a superposition of terms which factorize into a interface separation and translation part,
\begin{eqnarray}
V(W, \Delta x, \Delta y) &\simeq& \sum_j^N \delta(\kh\cdot\nh, \khj\cdot\nh) ] \times \nonumber \\
&& | c_{j, 0}|^2 f_{\kh\cdot\nh}(\Delta x, \Delta y) V_{\kh\cdot\nh}(W).
\end{eqnarray}


\subsection{\{100\} surfaces}

\begin{figure}
\begin{center}
\includegraphics[width=8cm]{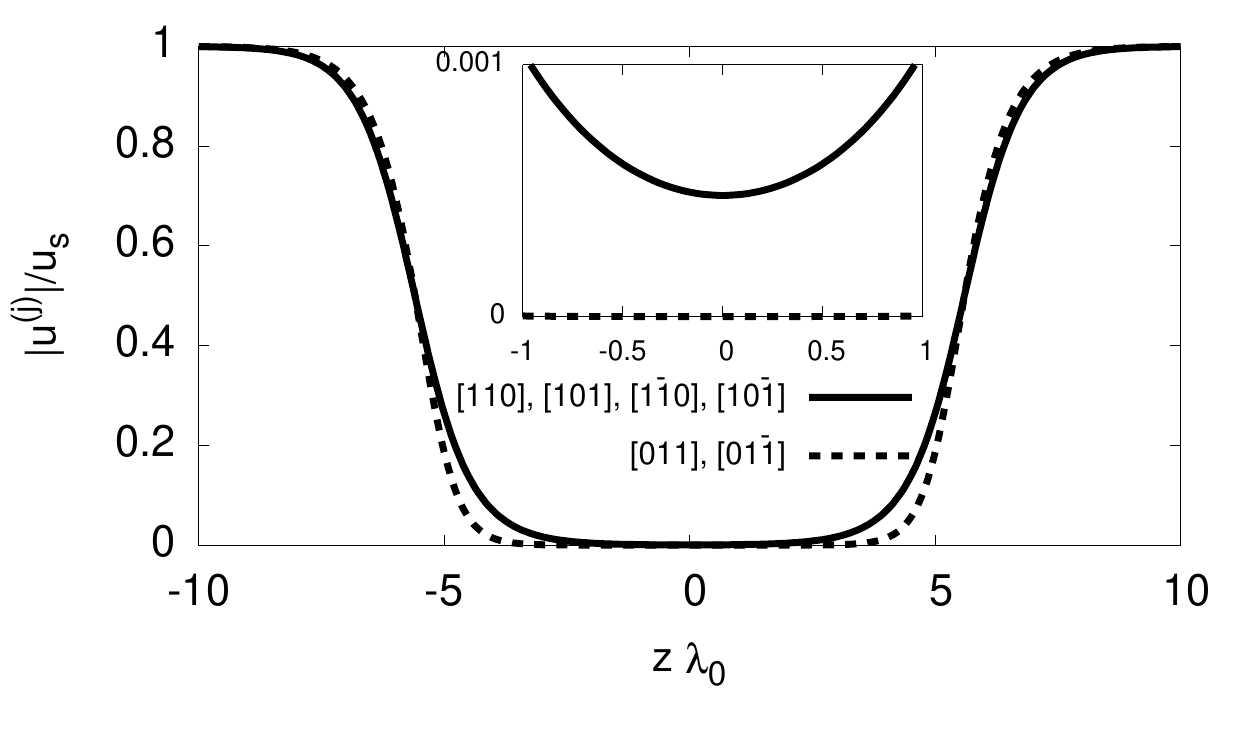}
\caption{Absolute value of the density wave amplitudes for (100) interfaces without lattice shift, $\psi_1^{(100)}=\psi_2^{(100)}=0$. 
The inset is a magnification around the origin, showing that the absolute value of the amplitudes varies smoothly there. The slaved fields decay quickly and do not show a visible overlap of incoming and outgoing waves. 
}
\label{shift::fig6}
\end{center}
\end{figure}

We consider two parallel (100) interfaces of bcc crystals as a first example.
Asymptotically, the interactions stem exclusively from the density waves with the slowest decay, i.e.~with the highest directional cosine $\khj\cdot\hat{n}$.
In this case, the principal reciprocal lattice vectors $[110], [101], [1\bar{1}0], [10\bar{1}]$ (and their inverses) have the same decay length, and the remaining, $[011], [01\bar{1}]$ (+ inverses) form a second group. 
All density waves within the same group have the same absolute value, but usually differ in phase;
notice that the amplitudes depend on the lattice shift.
This is shown in Fig.~\ref{shift::fig6}, where the absolute value of the density wave amplitudes is plotted as function of the position normal to the interfaces for a case without lattice shift.
In the solid, all amplitudes reach the same bulk value $u_s$ due to the crystallographic symmetries.

In agreement with the notation of (100) interfaces, we use a coordinate representation for $\nh^{(100)}=\zh^{(100)} = (1,0,0)$, and hence the tangential vectors have the coordinate representation  $\xh^{(100)}=(0,1,0)$ and $\yh^{(100)}=(0,0,1)$.
The translation is periodic with respect to shifts by one lattice unit $a=2\sqrt{2}\pi/q_0$ in each direction $\xh$ and $\yh$ (the factor $\sqrt{2}$ comes from the fact that the reciprocal lattice vectors point along the face diagonal of the bcc crystal).
We can therefore introduce rescaled coordinates $\psi_1^{(100)}= \Delta x\, q_0/\sqrt{2}$ and $\psi_2^{(100)}= \Delta y \,q_0/\sqrt{2}$;
all properties are then $2\pi$ periodic for this coordinate representation.

\begin{figure}
\begin{center}
\includegraphics[width=9cm]{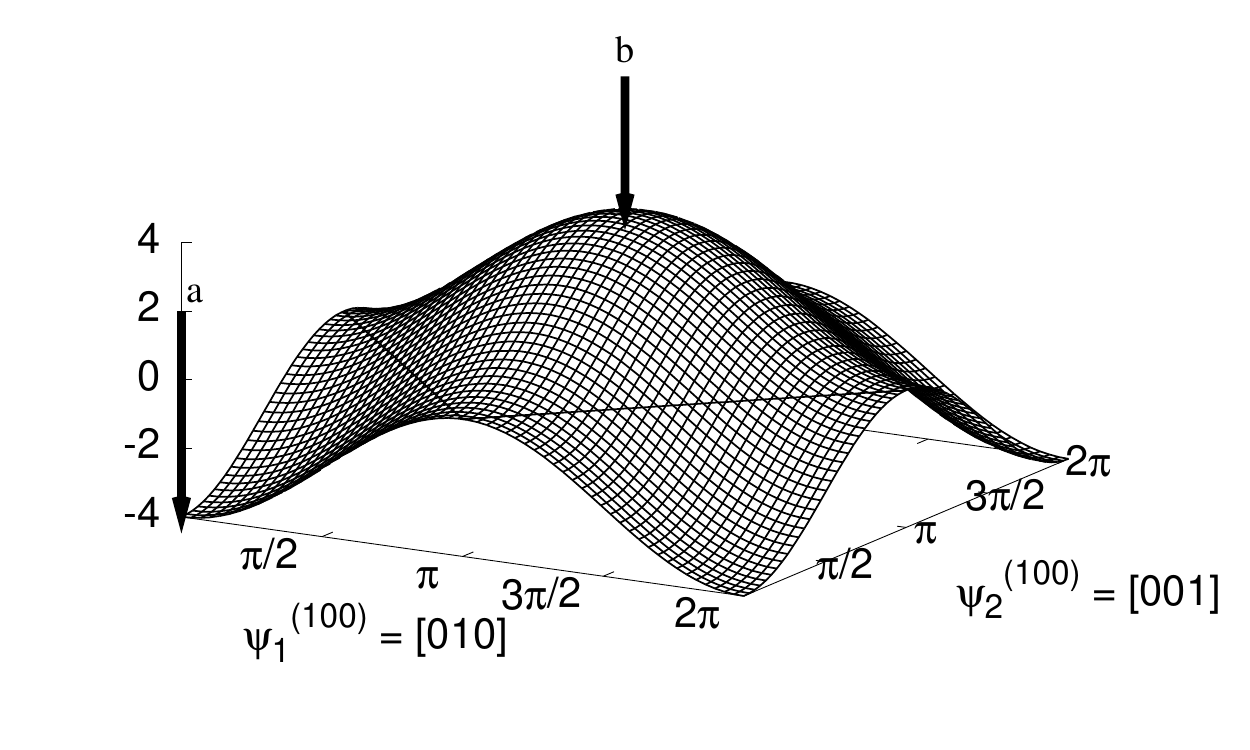}
\caption{Lateral dependence $f_{\kh\cdot\nh=1/\sqrt{2}}^{(100)}$ of the slowest decaying contribution to the disjoining potential of two parallel (100) surfaces. The interaction is repulsive for positive function value. The two arrows mark the most attractive (a) and most repulsive situation (b).
}
\label{shift::fig7}
\end{center}
\end{figure}

\begin{figure}
\begin{center}
\includegraphics[width=9cm]{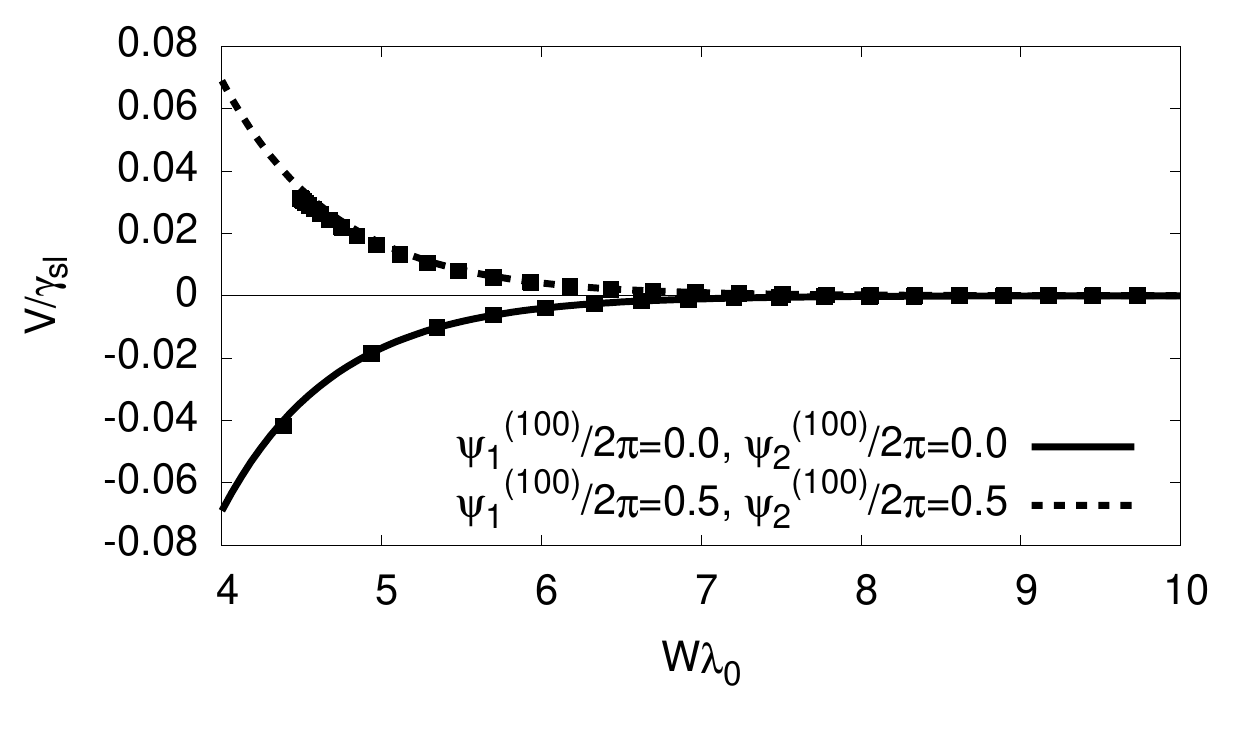}
\caption{The disjoining potential for the two cases (a) (solid curve) and (b) (dashed curve) of (100) interfaces. For each case, the squares show the result from the numerical simulation, the lines the asymptotic prediction, taking into account the slowest decaying density waves.
}
\label{shift::fig8}
\end{center}
\end{figure}

\begin{table}
\begin{center}
\begin{tabular}{c|c|c|c}
$\khj$ & $\khj\cdot\hat{n}$ & Phase shift $\phi_j$ & Matching constant $c_{j, 0}$ \\
\hline
$[110]$ & $1/\sqrt{2}$ & $\psi_1^{(100)}$ & 0.165 \\
$[101]$ & $1/\sqrt{2}$ & $\psi_2^{(100)}$ & 0.165 \\
$[1\bar{1}0]$ & $1/\sqrt{2}$ & $-\psi_1^{(100)}$ & 0.165 \\
$[10\bar{1}]$ & $1/\sqrt{2}$ & $-\psi_2^{(100)}$ & 0.165 \\
$[011]$ & 0 & (slaved) & (slaved) \\
$[01\bar{1}]$ & 0 & (slaved) & (slaved)
\end{tabular}
\caption{Matching and shift properties of the density wave amplitudes for (100) interfaces.}
\label{shift::table1}
\end{center}
\end{table}
For the (100) interfaces, the phase shifts of the fields are summarized in Table \ref{shift::table1} together with the previously determined matching constants, and we therefore obtain for the longest range exponentials
\begin{equation}
f_{\kh\cdot\nh=1/\sqrt{2}}^{(100)}(\psi_1^{(100)}, \psi_2^{(100)})=-2(\cos\psi_1^{(100)} +\cos\psi_2^{(100)}),
\end{equation}
which is plotted in Fig.~\ref{shift::fig7}.
The disjoining potential for the most attractive situation (matching lattices, (a)) and the most repulsive case (b) are shown in Fig.~\ref{shift::fig8}.
The predictions are compared to numerical calculations, which were obtained in a dynamical run at $T=T_M$.
This means that we set up a solid-liquid-solid ``sandwich'' structure, with a phase shift between the solid phases in the real space implementation.
Due to the overlap of the interface profiles we have an attractive or repulsive interaction between the interfaces, thus the configuration is not in full equilibrium.
During the time evolution we numerically compute the melt layer thickness $W$ and the energy $F$ according to Eq.~(\ref{fenAE}), without the correction term from the box operator.
The dependence $F(W)$ is then plotted and compared to the analytical predictions.
This method is approximative in the sense that the system is not in full equilibrium with $\partial \Aj/\partial t=0$.
A more precise approach is to balance the interaction with a thermal tilt $T\neq T_M$ and then to calculate the energy for the relaxed solution $F-F_T$;
this approach is used for the interaction of misoriented crystals, which are treated in Section \ref{misoriented}.
As long as the interaction is weak, both methods give the same results, and we have checked that the present results are robust.
Also, they agree very well with the analytical predictions for the asymptotic interaction.
For the special case that the disjoining potential has a minimum -- a case that we will encounter later --, the dynamical runs converge to this point, where the interaction energy therefore becomes exact.

\begin{figure}
\begin{center}
\includegraphics[width=8cm]{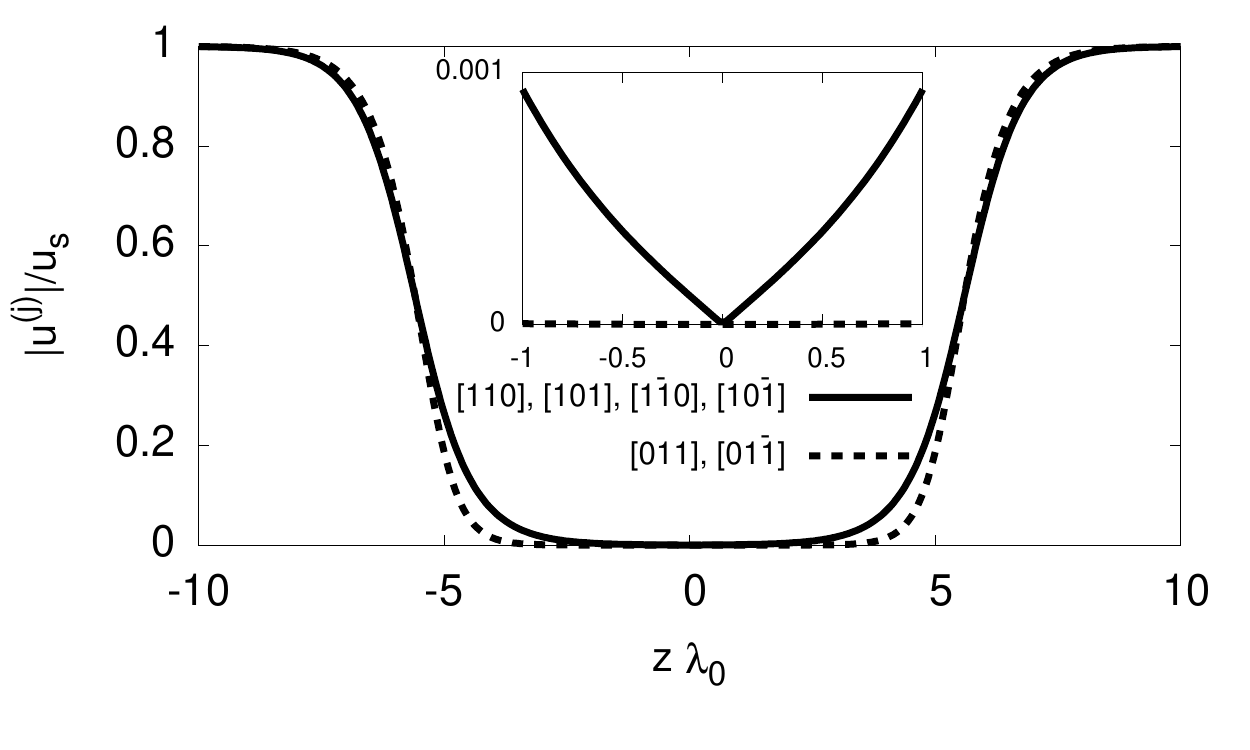}
\includegraphics[width=8cm]{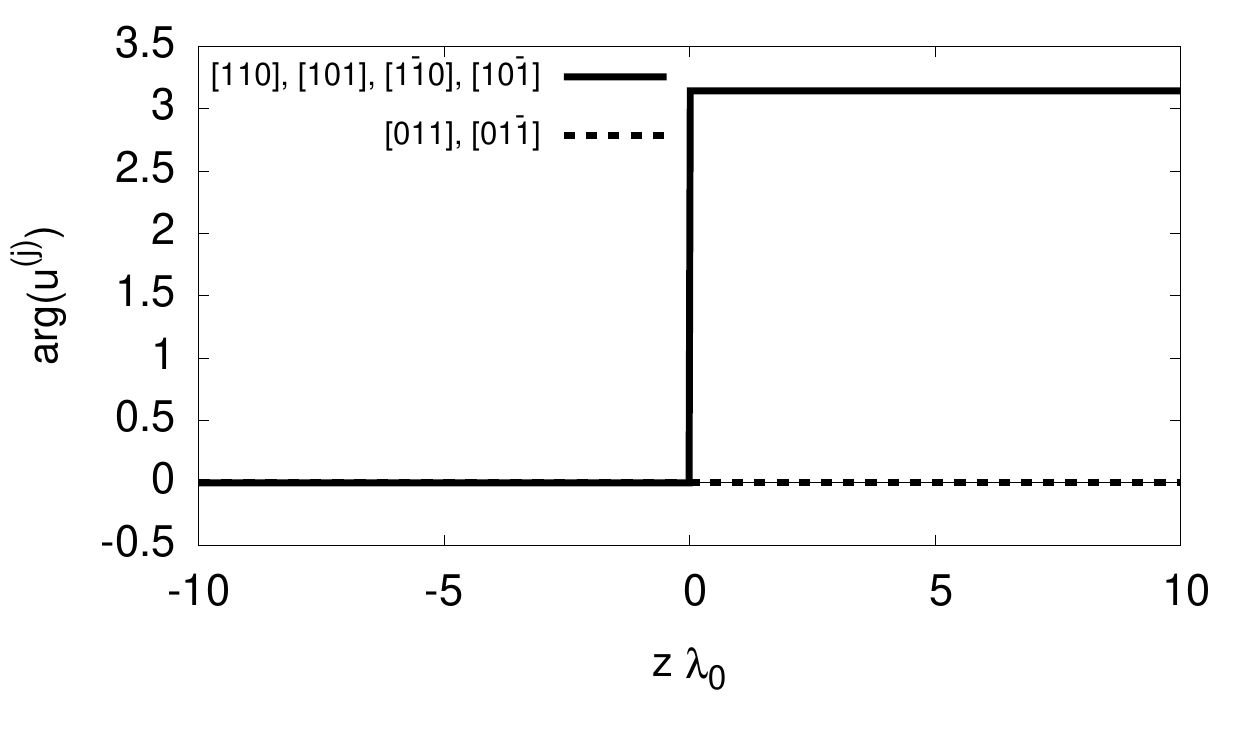}
\caption{Top: Absolute value of the density wave amplitudes for (100) interfaces for crystals that are shifted by half a lattice unit, $\psi_1^{(100)}=\psi_2^{(100)}=\pi$. 
The inset is a magnification around the origin, showing that the absolute value of the amplitudes has a cusp there.
The slaved fields decay quickly and do not show a visible overlap of incoming and outgoing waves.
Bottom: Corresponding phases of the amplitudes.
}
\label{shift::fig9}
\end{center}
\end{figure}

In the liquid region, the density wave amplitudes are given by the expression (\ref{single::eq2}).
It is instructive to look also at amplitude and phase separately.
With a real coefficient $c_{i, in}$ and $c_{j, out}=c_{j, in}\exp(i \phi_j)$ we obtain at $z=0$
\begin{equation}
| \uj |^2 = 2 c_{j,in}^2 (1+\cos\phi_j).
\end{equation}
For the special case $\phi_j=\pi$ (the most repulsive case) the amplitudes have a cusp there, and correspondingly the phase jumps discontinuously.
Notice, however, that this singular behavior appears only in the polar representation of the complex amplitudes;
in a complex sense they are smooth at $z=0$.
This behavior is visualized in Fig.~\ref{shift::fig9}.
We note that a cusp in the ``order parameter'' was introduced phenomenologically in Ref.~\onlinecite{LobkovskyWarren2002}.
Here it is a natural consequence of the description.

\subsection{\{110\} surfaces}

The situation immediately becomes more complex for the next example of (110) interfaces, thus $\nh^{(110)}=(1, 1, 0)/\sqrt{2}$.
The tangential vectors are here defined through the coordinate representation $\xh^{(110)}=(0, 0, 1)$ and $\yh^{(110)}=(1, -1, 0)/\sqrt{2}$, and the phase factors are $\psi_1^{(110)}=\Delta x\, q_0/\sqrt{2}$ and $\psi_2^{(110)}=\Delta y\, q_0/2$, to recover the $2\pi$ periodicity.
This means that the axes in the interface plane are stretched differently, and therefore the geometry loses its fourfold symmetry (it reduces to a $C_2$ symmetry), in agreement with the fact that in the interface plane the distances between the atoms are different for the two perpendicular directions.

Here, the density wave which corresponds to the principal reciprocal lattice vector $\khj=[110]$ has the longest range, since its crystallographic ordering extends the farthest into the melt, $\khj\cdot\nh=1$.
Therefore, the longest range interaction is mediated by this density wave.
Since it is a plane wave, it has no lateral dependence, which means that the disjoining potential does not depend on $\psi_1^{(110)}, \psi_2^{(110)}$, and this contribution to the interaction is always attractive.
This implies that at large distances we always find an attraction of crystals, irrespective of the lattice shift, i.e.
\begin{equation}
f_{\kh\cdot\nh=1}^{(110)} = -1.
\end{equation}

\begin{figure}
\begin{center}
\includegraphics[width=9cm]{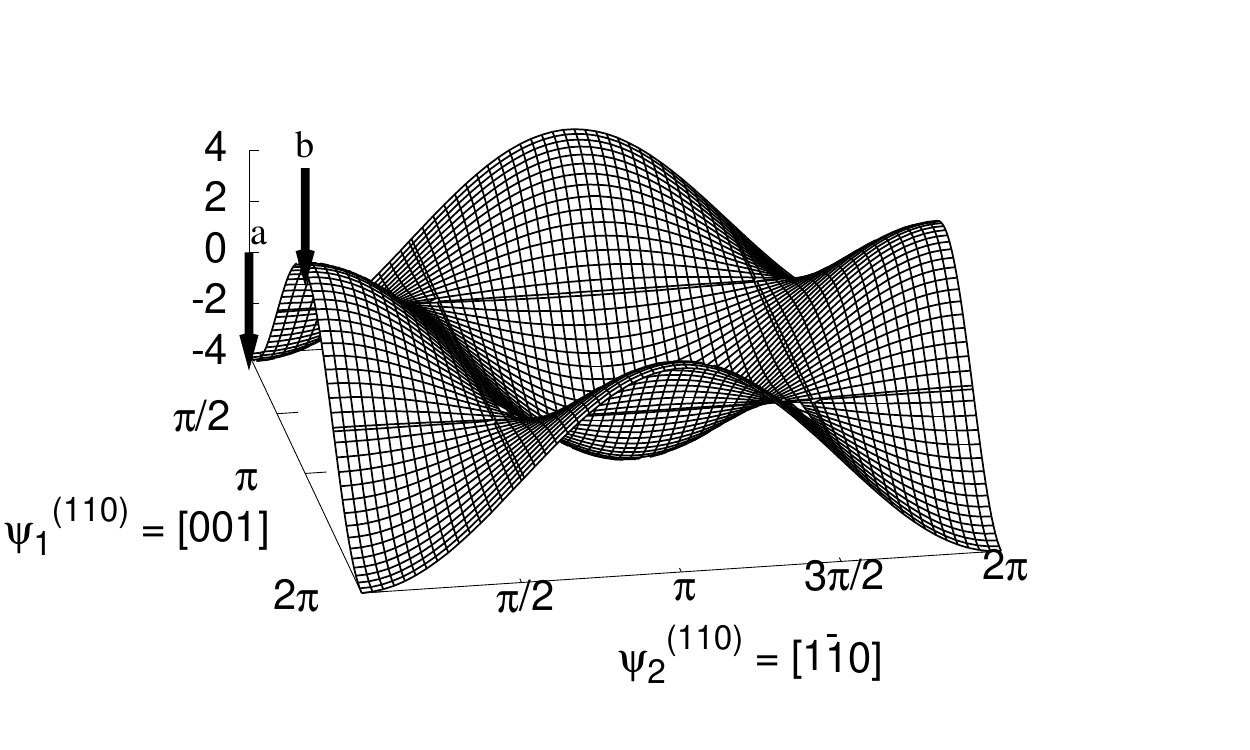}
\caption{Lateral dependence $f_{\kh\cdot\nh=1/2}^{(110)}$ of the second contribution to the disjoining potential of two parallel (110) surfaces. The interaction is repulsive for positive function value. The two arrows mark the most attractive (a) and most repulsive situation (b).
}
\label{shift::figa}
\end{center}
\end{figure}

As soon as the interfaces come closer to each other, the contribution from the next density waves becomes noticeable.
In this case, it comes from the density waves with reciprocal vectors [101], [011], $[10\bar{1}]$, $[01\bar{1}]$ (and their inverse vectors), which have all a directional cosine $\kh\cdot\nh=1/2$, thus the range of their contribution to the interaction is only half of the range of the leading term.
The total lateral dependence from this set of density waves,
\begin{equation}
f_{\kh\cdot\nh=1/2}^{(110)} =-4\cos\psi_1^{(110)}\cos\psi_2^{(110)},
\end{equation}
is shown in Fig.~\ref{shift::figa}.
It has attractive and repulsive regions:
If the crystals are perfectly aligned, $\psi^{(110)}_1=\psi^{(110)}_2=0$, the interaction is of course attractive, because the interfacial energy would vanish completely if the crystals merge.
For maximum mismatch, i.e.~if the crystals are shifted by half a lattice unit in one direction, the interaction has reached the strongest repulsive situation.
For a shift by half a lattice unit in both directions we recover again the attractive case, because then the crystallographic planes in the (110) surface match again.

\begin{figure}
\begin{center}
\includegraphics[width=8cm]{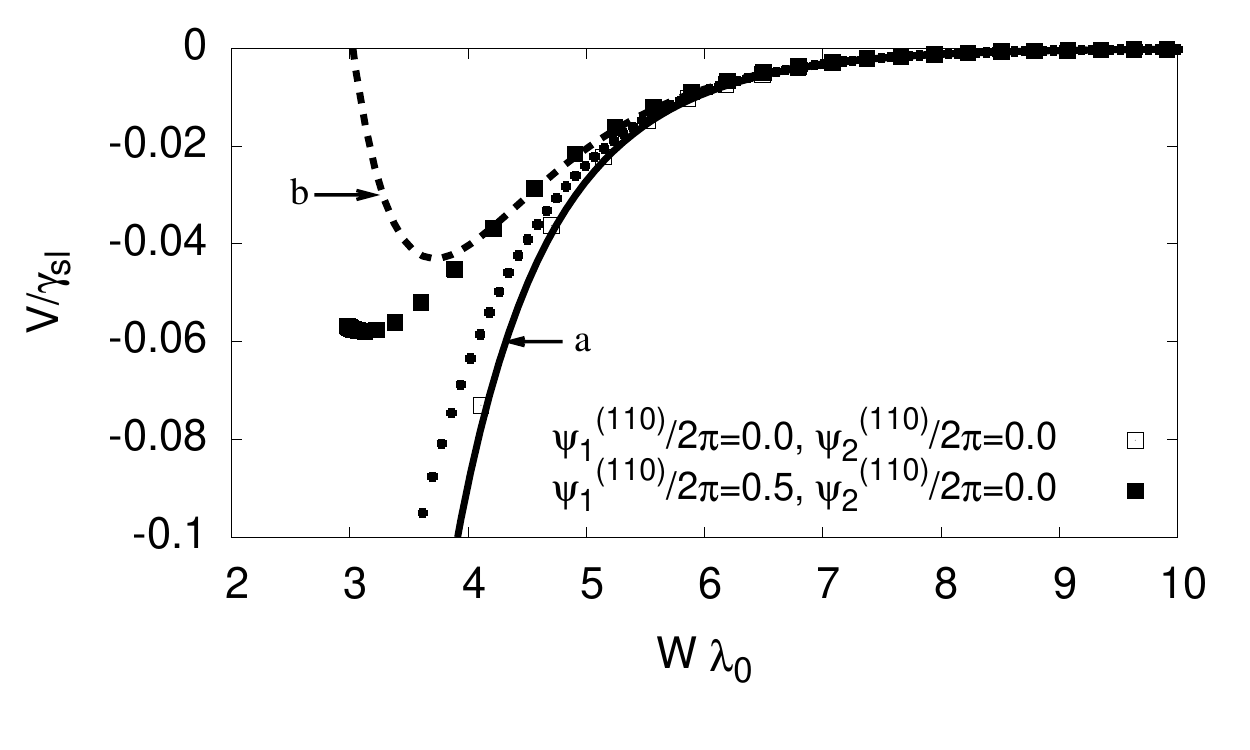}
\caption{Disjoining potential for (110) interfaces. At large distances, the interaction is always attractive and solely determined by the (110) density wave. The graphs shows numerical results for two different shifts together with the analytical predictions: The dotted curve takes into account only the longest-range exponential, the solid (a) and dashed curve (b) also the corrections due to faster decaying density waves.
}
\label{shift::fig10}
\end{center}
\end{figure}

\begin{figure}
\begin{center}
\includegraphics[width=8cm]{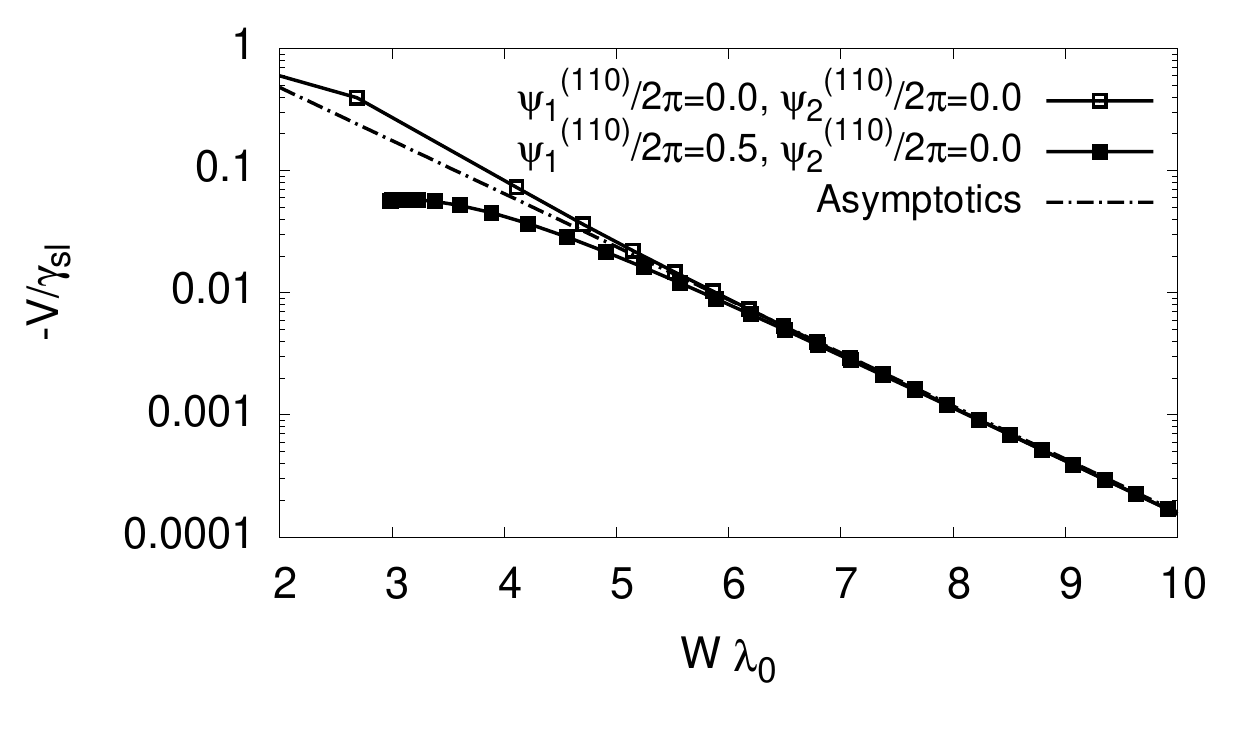}
\caption{Asymptotics of the disjoining potential for (110) interfaces. At large distances, the interaction is always attractive and solely determined by the (110) density wave. The graphs shows numerical results for two different shifts together with the analytical prediction (only the contribution from the slowest decaying exponential).
}
\label{shift::fig11}
\end{center}
\end{figure}

\begin{table}
\begin{center}
\begin{tabular}{c|c|c|c}
$\khj$ & $\khj\cdot\hat{n}$ & Phase shift $\phi_j$ & Matching constant $c_{j, 0}$ \\
\hline
$[110]$ & $1$ & $0$ & 0.116 \\
$[101]$ & $1/2$ & $\psi_1^{(110)}-\psi_2^{(110)}$ & 0.372 \\
$[011]$ & $1/2$ & $\psi_1^{(110)}+\psi_2^{(110)}$ & 0.372 \\
$[10\bar{1}]$ & $1/2$ & $-\psi_1^{(110)}-\psi_2^{(110)}$ & 0.372 \\
$[01\bar{1}]$ & $1/2$ & $-\psi_1^{(110)}+\psi_2^{(110)}$ & 0.372 \\
$[1\bar{1}0]$ & 0 & (slaved) & (slaved) \\
\end{tabular}
\caption{Matching and shift properties of the density wave amplitudes for (110) interfaces.}
\label{shift::table2}
\end{center}
\end{table}
The relevant data for the calculation of the asymptotic disjoining potential is summarized in Table \ref{shift::table2}, and the potential is plotted in Fig.~\ref{shift::fig10} for the case of no misfit (a) and the most repulsive case (b) with $\psi_1^{(110)}=\pi$ and $\psi_2^{(110)}=0$, as illustrated by the arrows in Fig.~\ref{shift::figa}.
As mentioned before, we have a purely attractive behavior at large distances independent of the lattice translation, and the agreement with the analytical prediction is confirmed in the logarithmic plot Fig.~\ref{shift::fig11}.
At shorter distances, the numerically calculated disjoining potential deviates from the analytical prediction from the slowest decaying density waves only (dotted lines in Fig.~\ref{shift::fig10}), and the inclusion of the next terms leads to a significantly better agreement (solid and dashed line), and we observe the distinction between the attractive and repulsive cases.
Only at short distances is the interaction strongly affected by nonlinear contributions.
In particular, we observe a stable minimum in the disjoining potential for the most repulsive case, because a hard core repulsion due to elastic deformations prevents a full merging of the interfaces.

\subsection{\{310\} surfaces}

\begin{figure}
\begin{center}
\includegraphics[width=10cm]{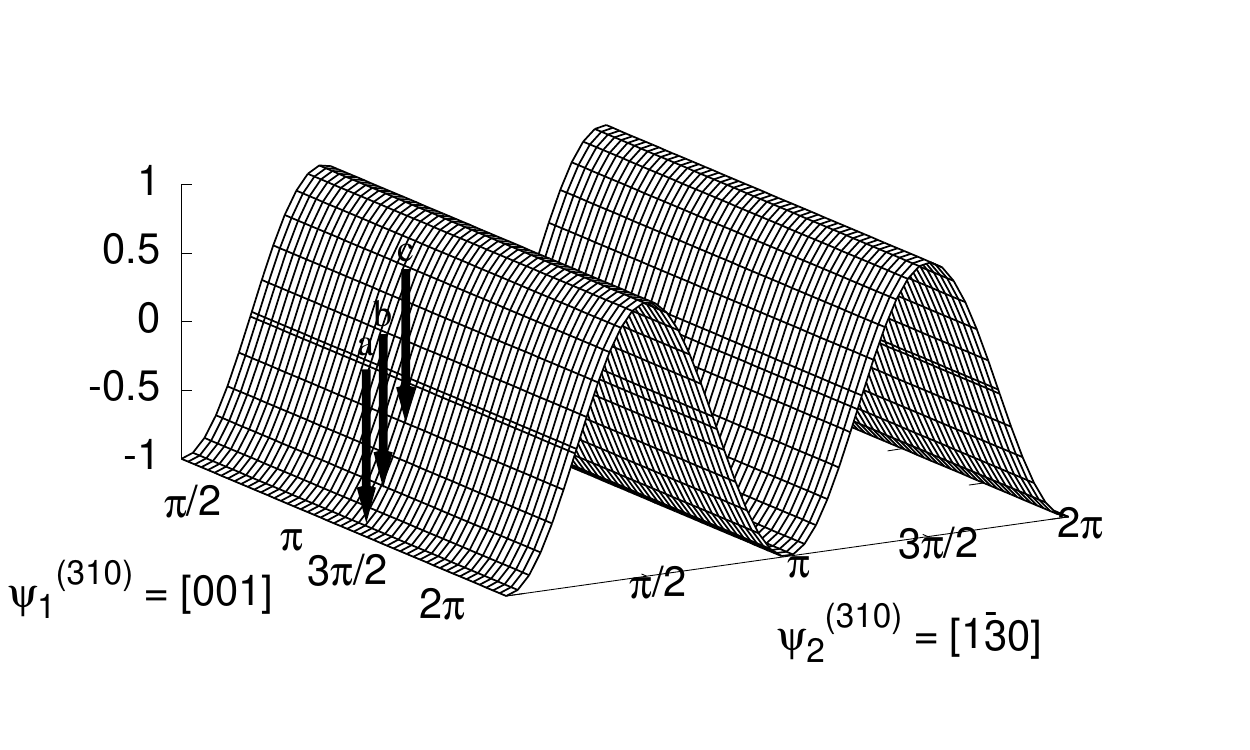}
\end{center}
\caption{Lateral dependence of the longest range contribution of the disjoining potential for the (310) interfaces. The potential is attractive in regions where the value is negative. In particular, this contribution to the interaction does not depend on the translation in the [001] direction.
}
\label{shift::fig3}
\end{figure}

\begin{figure}
\begin{center}
\includegraphics[width=10cm]{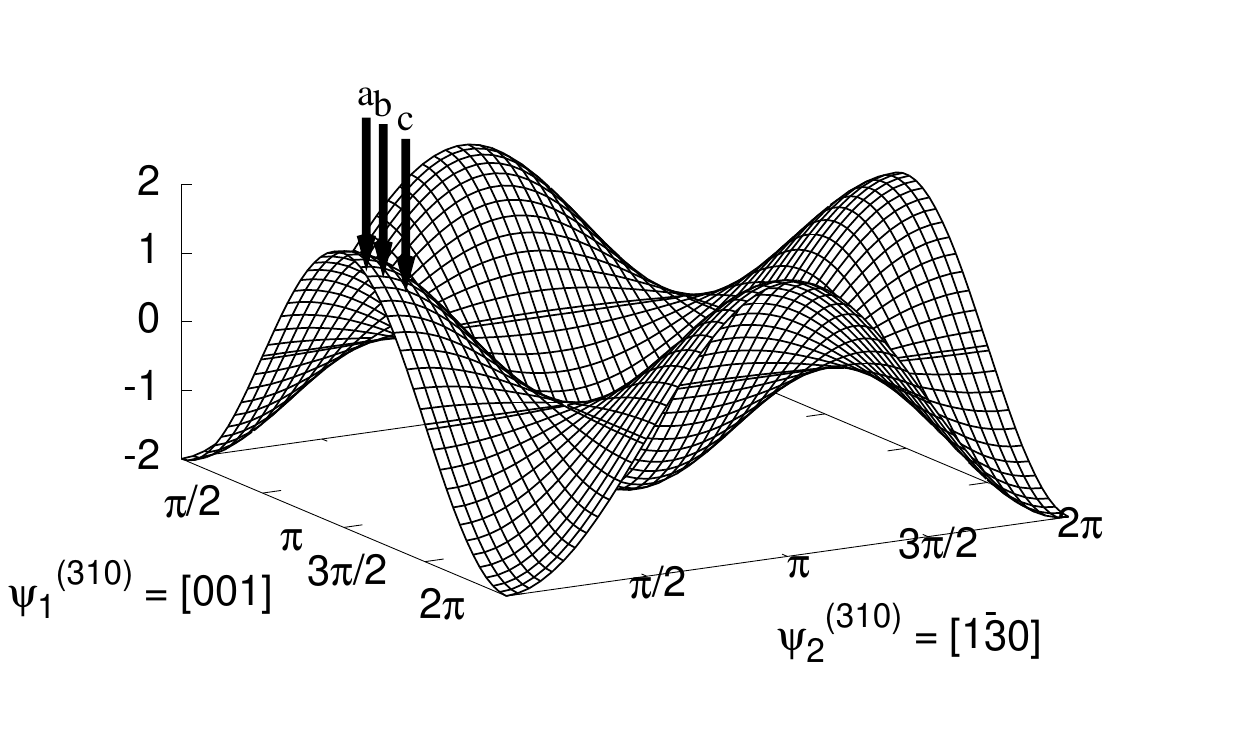}
\end{center}
\caption{Lateral dependence of the second longest range contribution of the disjoining potential for the (310) interface normal direction. The potential is attractive in regions where the value is negative.
}
\label{shift::fig4}
\end{figure}

As a last example we investigate (310) interfaces, where we receive nontrivial contributions from the first and the second exponentials.
With $\nh^{(310)}=(3,1,0)/\sqrt{10}$ we use tangential vectors $\xh^{(310)}=(0,0,1)$ and $\yh^{(310)}=(1, -3, 0)/\sqrt{10}$.
The $2\pi$ periodic in-plane coordinates are $\psi_1^{(310)} = \Delta x\, q_0/\sqrt{2}$ and $\psi_2^{(310)} = \Delta y\, q_0/\sqrt{20}$, and all data is summarized in Table \ref{shift::table3}.
The longest range density wave is $u_{110}$, and the lateral dependence of the disjoining potential
\begin{equation}
f_{\kh\cdot\nh=2/\sqrt{5}}^{(310)} = -\cos 2\psi_2^{(310)}
\end{equation}
is shown in Fig.~\ref{shift::fig3}.
Two fields, $[101]$ and $[10\bar{1}]$ (plus inverse vectors), contribute to the next exponential, therefore the lateral dependence of this term is
\begin{eqnarray}
f_{\kh\cdot\nh = 3/2\sqrt{2}}^{(310)} &=& -\cos(\psi_1^{(310)}+\psi_2^{(310)}) - \cos(\psi_1^{(310)}-\psi_2^{(310)}) \nonumber \\
&=& -2\cos\psi_1^{(310)}\cos\psi_2^{(310)}, 
\end{eqnarray}
see Fig.~\ref{shift::fig4}.
\begin{table}
\begin{center}
\begin{tabular}{c|c|c|c}
$\khj$ & $\khj\cdot\hat{n}$ & Phase shift $\phi_j$ & Matching constant $c_{j, 0}$ \\
\hline 
$[110]$ & $2/\sqrt{5}\approx 0.89$ & $-2\psi_2^{(310)}$ & 0.128 \\
$[101]$ & $3/2\sqrt{2}\approx 0.67 $ & $\psi_1^{(310)}+\psi_2^{(310)}$ & 0.18\\
$[10\bar{1}]$ & $3/2\sqrt{2}\approx 0.67$ & $\psi_2^{(310)}-\psi_1^{(310)}$ & 0.18 \\
$[1\bar{1}0]$ & $1/\sqrt{5} \approx 0.44$ & $4\psi_2^{(310)}$ & 0.425 \\
$[011]$ & $1/2\sqrt{5} \approx 0.22$ & (slaved by cubic) & (slaved by cubic) \\
$[01\bar{1}]$ & $1/2\sqrt{5} \approx 0.22$ & (slaved by cubic) & (slaved by cubic)
\end{tabular}
\caption{Phase shifts and matching constants for the (310) interface normal for the non-slaved fields. Notice that the fields with the highest scalar product $\khj\cdot\hat{n}$ have the longest range.}
\label{shift::table3}
\end{center}
\end{table}

\begin{figure}
\begin{center}
\includegraphics[width=9cm]{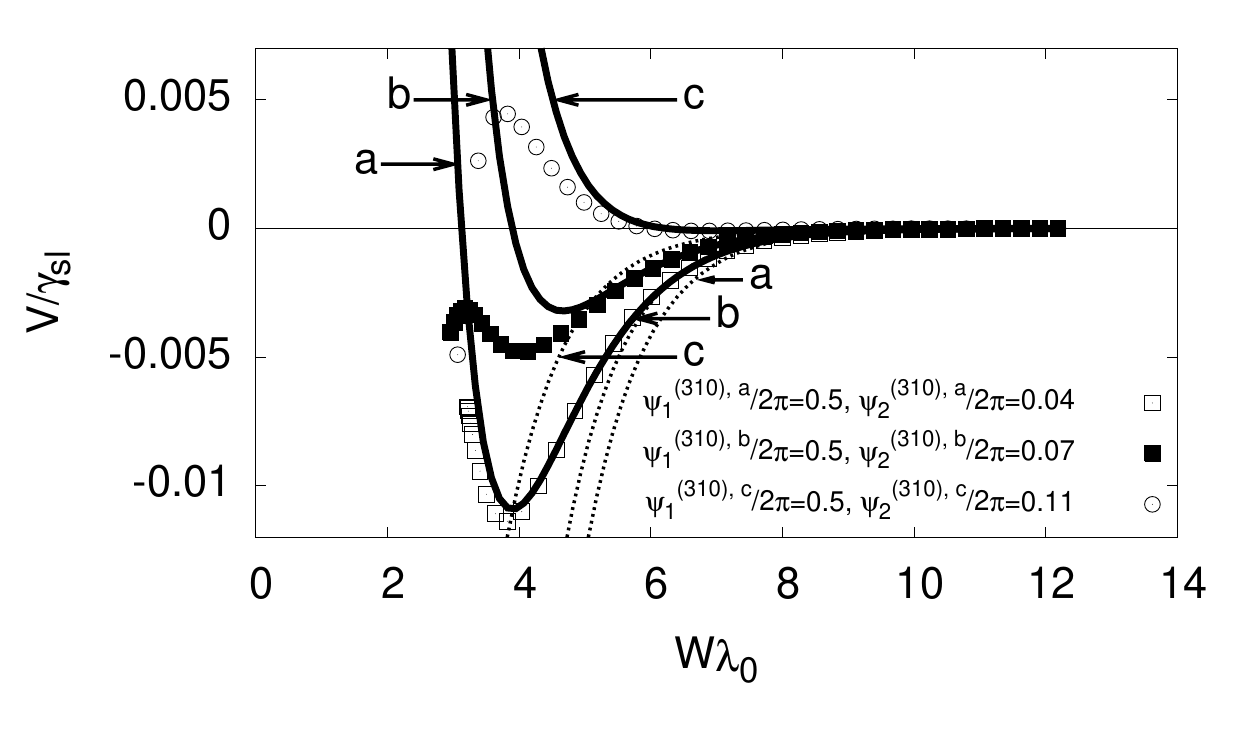}
\caption{The disjoining potential for (310) interfaces for the three cases a, b, c, as explained in the text.
For each case, the isolated points show the result from the numerical simulation, the dotted lines the asymptotic prediction, taking into account only the slowest decaying density wave [110], and the solid line the analytical predictions, using the first and second exponentials. The potential is here in all cases attractive at large distances and repulsive for small separations $W$.
}
\label{shift::fig5}
\end{center}
\end{figure}

\begin{figure}
\begin{center}
\includegraphics[width=9cm]{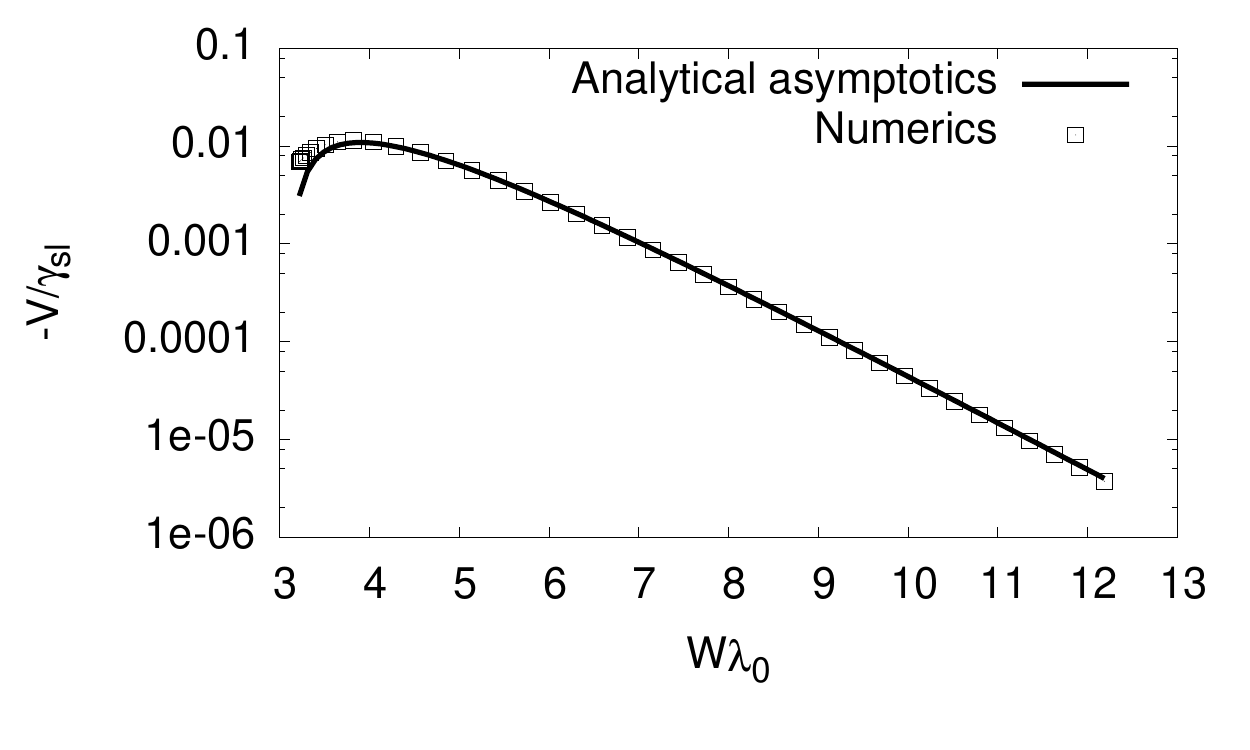}
\caption{Logarithmic plot of the asymptotics of the disjoining potential for $\psi_1^{(310), a}=\pi$ and $\psi_2^{(310), a} = 0.04\cdot 2\pi$ for (310) interfaces.
The numerical results are compared against the analytically determined asymptotic behavior including the two slowest decaying exponentials.}
\label{shift::fig2}
\end{center}
\end{figure}

We investigate in particular three different shifts, all with $\psi_1^{(310), (a, b, c)}=\pi$ and $\psi_2^{(310), a}=0.04\cdot 2\pi$, $\psi_2^{(310), b}=0.07\cdot 2\pi$ and $\psi_2^{(310), c}=0.11\cdot 2\pi$.
For this inclination, both the longest range term and the next term have -- depending on the mismatch -- attractive and repulsive regions.
The three scenarios are indicated by the arrows in Fig.~\ref{shift::fig3} for the lateral dependence of the slowest decaying exponential and also in Fig.~\ref{shift::fig4} for the next exponential.
Apparently, for the sample cases (a), (b), and (c) the slowest decaying exponential is always attractive, whereas the second is repulsive.
From (a) to (c) the strength of the first exponential becomes smaller, and therefore we see a crossover from a long-range attraction to an intermediate repulsion.
This prediction is confirmed by the numerical results in Fig.~\ref{shift::fig5}.
Fig.~\ref{shift::fig2} shows the asymptotics of the disjoining potential for $\psi_1^{(310), a}=\pi$ and $\psi_2^{(310), a} = 0.04\cdot 2\pi$ and the comparison with the analytical prediction.

\subsection{Box operator corrections}

Let us briefly discuss the influence of the correction terms of the box operator, which have been neglected in the discussion so far.
With the previous knowledge that the additional terms are small, it is immediately transparent that the results can be modified only slightly, and therefore the above simplified picture remains valid.
Nevertheless, the analysis can also be formally performed in this more complicated case, and this is outlined here.

The presence of the box operator leads to the following modifications:
First, the amplitudes in the linearized region become a superposition of four exponential solutions instead of only two.
Determining the disjoining potential now requires identifying matching pairs of incoming and outgoing waves in the sense of a conservation law.
Two of the exponentials are strongly suppressed, since they show relatively fast oscillations.
Second, the concept of the Hamiltonian as in classical mechanics is only applicable if the free energy density contains only first order derivatives (the kinetic term).
The box operator, however, introduces higher order derivatives, and therefore this concept has to be generalized.

A generalized conservation law, which is valid also for misoriented grain boundaries, where the box operator is essential, is derived in Appendix \ref{gencons}, and here we need only the special case that all fields depend only on the coordinate normal to the grain boundary.
(In the general case, the Hamiltonian is an integral expression along the grain boundary plane, which reflects the fact that the interaction forces can vary spatially and have to be averaged to get the net force.)
Here, the interaction is homogeneous, and therefore the following expression becomes a conserved quantity:
\begin{eqnarray}
H &=&  \sum_{j=1}^{N/2} \Big( \pj \duj + \pjs \dujs + \rj \dduj + \rjs \ddujs \nonumber \\
&&- \drj \duj - \drjs \dujs \Big) - f,
\end{eqnarray}
where $f$ is the free-energy density.
It corresponds to a generalized Legendre transformation, with ``momenta''
\begin{equation}
\pj := \frac{\partial f}{\partial \duj}, \qquad \rj := \frac{\partial f}{\partial \dduj},
\end{equation}
where we treat $\uj$ and the complex conjugate $\ujs$ as independent functions.
The Hamiltonian is conserved, i.e. $\dot{H}=0$.
From the linearized solution (\ref{singleint::eq1}) we obtain after some straightforward but tedious algebraic manipulations
\begin{equation}
H = -\frac{2n_0 k_B T}{S(q_0)} \sum_j^{N/2} \left( \cjams\cjap + \cjaps\cjam + \cjbms\cjbp + \cjbps\cjbm \right)
\end{equation}
for the value of the Hamiltonian, calculated in the liquid up to second order.

Similar to before, the prefactors of the exponentials in the linearized solution decay with the melt layer thickness, and we have
\begin{eqnarray}
\cjam &=& \cjamn \exp(\lambda_{j, a}^- W/2), \\
\cjap &=& \cjapn \exp(-\lambda_{j, a}^+ W/2), \\
\cjbm &=& \cjbmn \exp(\lambda_{j, b}^- W/2), \\
\cjbp &=& \cjbpn \exp(-\lambda_{j, b}^+ W/2).
\end{eqnarray}
Therefore, the disjoining potential becomes
\begin{eqnarray}
V(W) &=& -\frac{2n_0 k_B T}{S(q_0)} \sum_j^{N/2} \Big[ \frac{\cjamns\cjapn}{\lambda_{j, a}^+} \exp(-\lambda_{j,a}^+ W) \nonumber \\
&+& \frac{\cjbmns\cjbpn}{\lambda_{j, b}^+} \exp(-\lambda_{j,b}^+ W)\Big] + c.c.
\end{eqnarray}
Again, the prefactors acquire a complex factor if the crystals are translated against each other.
As mentioned before, the prefactors $\cjbmn$ and $\cjbpn$ are small and can be neglected.
For rough interfaces ($\epsilon\to 0$) we recover the above expression (\ref{shift::eq12}) for the disjoining potential.



\section{Interaction between misoriented grains}
\label{misoriented}

An analysis for the long range interaction as for the shifted crystals is not possible if a misorientation is involved, since the problem is not one-dimensional anymore.
However, the range of the interactions can still be understood using similar arguments, and the central outcome is that they are significantly shorter ranged.
We consider here the case of a tilt grain boundary, to illustrate the basic idea.
As had been discussed in detail in Ref.~\onlinecite{SpaKar09} the presence of a lattice rotation makes the use of the full box operator mandatory, and still the description is only valid for small misorientations.

Let us assume that the left grain ($z\to-\infty$, characterized by subscript $-$) is rotated by $\Phi_-$ with respect to the reference orientation of the RLVs, whereas the right grain ($z\to\infty$, subscript $+$) is rotated by $\Phi_+$.
Still the interfaces are assumed to be planar and normal to the $z$ axis.

First, the left grain has amplitudes
\begin{equation}
\uj_- = u_s \exp [i\kvjd \rotM(\Phi_-)\vec{r}]
\end{equation}
in the bulk, where the dagger denotes transposition and $\rotM(\Phi)=\rotR(\Phi)-\rotI$ with the identity matrix $\rotI$ and the rotation matrix $\rotR(\Phi)$,
\begin{equation}
\rotR(\Phi) = \left(
\begin{array}{cc}
\cos\Phi & \sin\Phi \\
-\sin\Phi & \cos\Phi
\end{array}
\right).
\end{equation}
Similarly, for the right grain
\begin{equation}
\uj_+- = u_s \exp [i\kvjd \rotM(\Phi_+)\vec{r}].
\end{equation}
This suggests looking for solutions in the liquid region of the structure
\begin{eqnarray} \label{misoriented::eq1}
\uj &=& c_{j, a}^- \exp [i\kvjd \rotM(\Phi_-)\vec{r}] \exp\left(\lambda_{j, a}^-(\Phi_-) z\right) \nonumber \\
&+& c_{j, b}^- \exp [i\kvjd \rotM(\Phi_-)\vec{r}] \exp\left(\lambda_{j, b}^-(\Phi_-) z\right) \nonumber \\
&+& c_{j, a}^+ \exp [i\kvjd \rotM(\Phi_+)\vec{r}] \exp\left(\lambda_{j, a}^+(\Phi_+) z\right) \nonumber \\
&+& c_{j, b}^+ \exp [i\kvjd \rotM(\Phi_+)\vec{r}] \exp\left(\lambda_{j, b}^+(\Phi_+) z\right),
\end{eqnarray}
in analogy to Eq.~(\ref{singleint::eq1}), with $\Re(\lambda_{j, a}^-)<0$, $\Re(\lambda_{j, b}^-)<0$, $\Re(\lambda_{j, a}^+)>0$, $\Re(\lambda_{j, b}^+)>0$.
The ranges $\lambda_{j, a/b}^{\pm}$ are computed from the linearized equilibrium condition (\ref{singleint::eq1a}) with the help of the rotation theorem
\[
\Box_j^2\left[ f(\rv) \exp(i\kvjd\rotM\rv) \right] = \exp(i\kvjd\rotM\rv) \Box_{j, +}^2 f(\rv)  ,
\]
for any function $f(\rv)$ and
\begin{equation}
\Box_{j, +} = \hat{k}^{(j)}_+\cdot\nabla - \frac{i}{2q_0} \nabla^2
\end{equation}
with the rotated reciprocal vectors $\hat{k}_+^{(j)} = \rotR^\dagger \khj$, see Ref.~\onlinecite{SpaKar09} for details.
It turns out that the decay ranges are given by the same expressions as for the shifted crystals, Eqs.~(\ref{boxrange1})-(\ref{boxrange4}), but the reciprocal vectors have to be rotated here appropriately in the $\khj\cdot\hat{n}$ term.

Inserting these expressions into the generalized conservation law (\ref{gencons::eq5}) using only quadratic terms gives $H=0$ in disagreement with the tilt $H=-\Delta f$.
This shows that the longest range interaction is not mediated by the quadratic terms in the functional but stems from the higher order nonlinearities.
Since their contribution vanished quickly in the melt phase, it is intuitively clear, that the interaction range for misoriented grains is shorter than for shifted crystals.

One can interpret this statement also in a physical way:
For two misoriented grains the normal shift between lattice planes of the two crystals varies along the grain boundary, and therefore the interface alternatingly consists of region, where the atomic planes match and where they are out of phase.
This leads to alternations of attractive and repulsive regions along the grain boundary.
Since the strength and size of attractive and repulsive regions is equal at quadratic order, their contributions cancel each other in the total interaction energy between the grains.
Thus, only shorter range higher order terms can be responsible for the disjoining potential here.

To understand the range of the remaining interaction further, one can continue to employ the conservation law (\ref{gencons::eq5}).
The next step is to assume that the interaction stems from the cubic terms in the functional.
They can appear in two different ways:
First, products of two incoming and one outgoing waves or one incoming and two outgoing waves from expression (\ref{misoriented::eq1}).
Second, the cubic nonlinearities (which appear as quadratic terms in the equilibrium conditions) generate perturbations of the basic solution (\ref{misoriented::eq1}).
The structure of these perturbations $\delta\uj$ would be again be a product of two density waves, and a product of the type $\uj\delta\uj$ would then have then same structure of a product of three density waves.
However, such a product would contain a terms like $\exp\left[(\lambda_1+\lambda_2+\lambda_3) z\right]$ with three decay lengths $\lambda_1$, $\lambda_2$, $\lambda_3$, which would appear in the conservation law.
Since it has a nontrivial $z$ dependence, all these terms in the end have to cancel, since by the conservation law the Hamiltonian must be $z$ independent.
Therefore, also the cubic terms cannot contribute to the long-range interaction.

\begin{figure}
\begin{center}
\includegraphics[width=8cm]{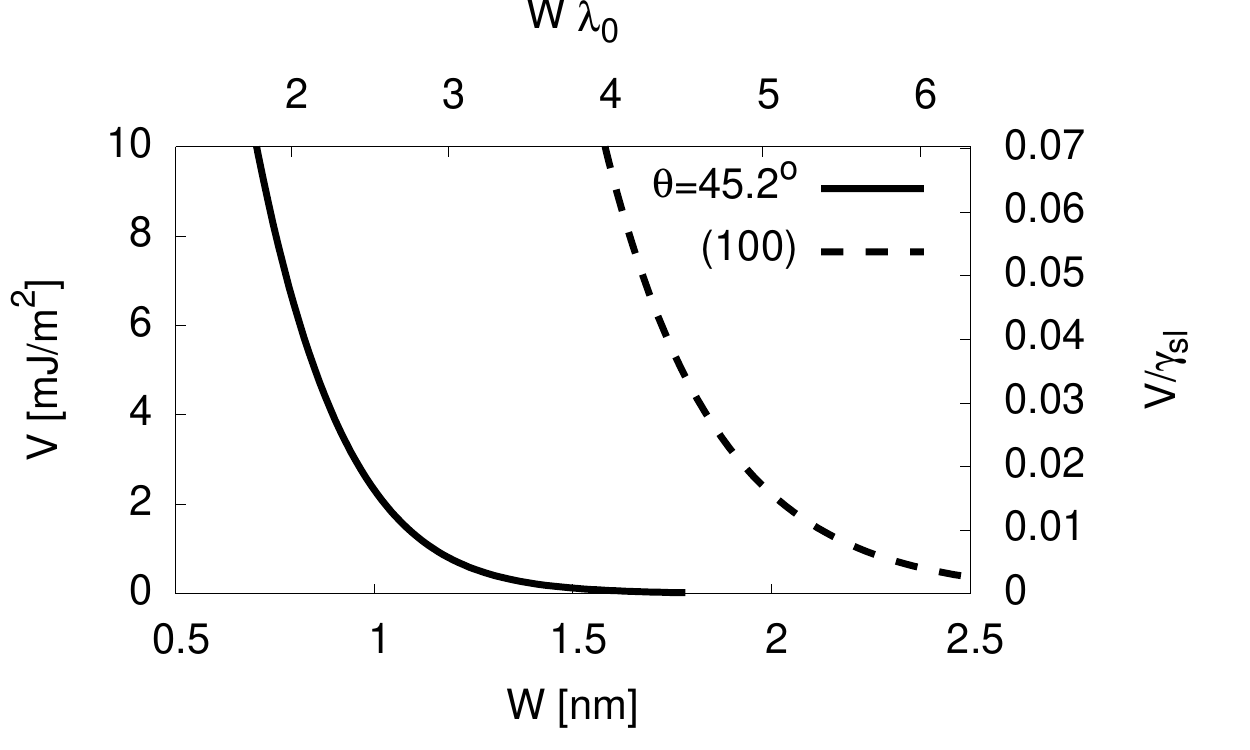}
\caption{Comparison of the interaction decay for (100) shifted crystals and a (100) symmetric tilt with $2\theta=45.2^\circ$ for $\delta$-iron.}
\label{misoriented::fig1}
\end{center}
\end{figure}

At quartic order, terms like $\exp\left[(\lambda_1+\lambda_2+\lambda_3+\lambda_4) z\right]$ can again be $z$ independent in the Hamiltonian, and they do not cancel, thus they finally balance the temperature term $\Delta f$.
The interaction range is therefore set by the sum of two $\lambda$ values, and decays therefore twice faster as for shifted crystals.
This is also confirmed by numerical investigations.
Fig.~\ref{misoriented::fig1} shows the cases of the shifted crystals in comparison to a repulsive grain boundary.
The shifted crystals case depicts the maximum repulsion for (100) interfaces, as discussed before in Fig.~\ref{shift::fig8} (curve b), again without the correction term of the box operator.
In comparison, the disjoining potential for the repulsive grain boundary with a misorientation of $2\theta=45.2^\circ$ decays significantly faster.
For these simulations we used $\epsilon=0.0923$ and also took into account the box operator correction, since otherwise rotated crystals would melt spuriously at $T=T_M$.
For the mapping to physical data we used the parameters for bcc $\alpha$-iron, which were previously determined in Ref.~\onlinecite{Wuetal06}.

\begin{figure}
\begin{center}
\includegraphics[width=9cm]{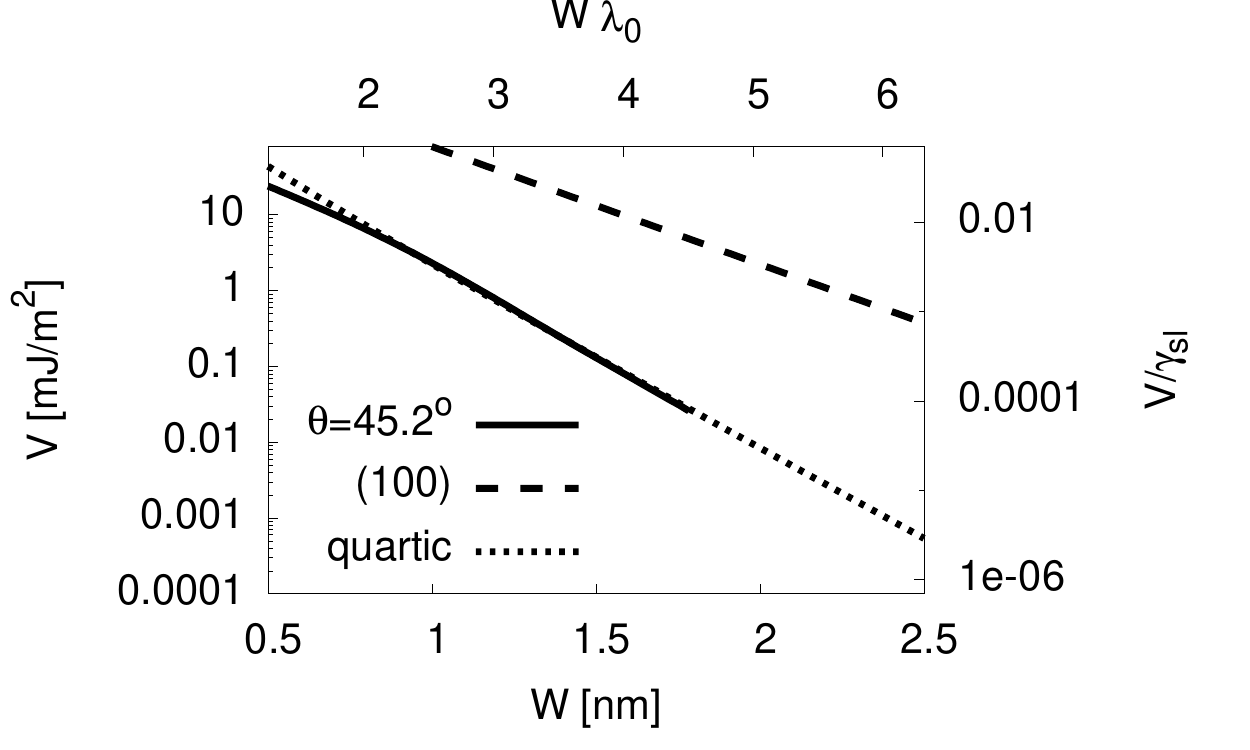}
\caption{Comparison of the interaction decay for (100) shifted crystals and a (100) symmetric tilt with $2\theta=45.2^\circ$ for $\delta$-iron.
The decay range of the quartic asymptotics is calculated without including the box operator correction term, which is negligible here.}
\label{misoriented::fig2}
\end{center}
\end{figure}

Fig.~\ref{misoriented::fig2} shows the same data on a logarithmic scale, together with the analytical prediction of the slowest decaying quartic interaction term, which stems from the $[110]$ density waves;
here we note that for the prediction of the decay range also the rotation of the interface normal by $\theta$ has to be taken into account in expression (\ref{single::eq3}).
The prefactor of the interaction energy is matched to the computed disjoining potential.
These numerical results confirm that the interaction of misoriented crystals is mediated by quartic terms in the framework of this model, and therefore the interactions of wet grain boundaries is very short ranged.

\section{Discussion and summary}

We have calculated the interaction between solid-liquid interfaces based on amplitude equations, which are derived from PFC or density functional theory.
In the framework of this model, the tail of the structural interaction can be calculated fully analytically for grains which are not misoriented but only differ by a lateral translation.
It is short-ranged and decays exponentially with the grain separation.
Depending on the lattice mismatch we find that the interaction is either attractive or repulsive.
It is most attractive if the lattice planes are fully aligned, such that complete freezing would remove any interface between the crystals.
In the opposite extreme case, that the grains are shifted against each other, such that a strong mismatch appears when the liquid layer disappears, leading to strong elastic deformations, the interaction is repulsive.
The entire interaction is a superposition of the contributions of the different density waves, and the longest range fields at a solid-liquid interface, i.e.~those density waves which extend the crystalline ordering furthest into the melt, also give the longest-range contribution to the solid-melt interface interaction.
The range of this interface interaction is given by Eq.~(\ref{single::eq3}) for the individual density waves.
This analytical expression also shows that the range of the interaction is determined by scattering properties of the melt phase and the relative orientation of the density wave vector to the interface normal.

The case of purely translated grains describes unstable configurations or repulsive interactions, since strong lateral forces will force the system back to configurations with aligned crystallographic planes.
Nevertheless, the results shall be relevant for the understanding of the merging of dendrite sidearms from the same grain, where due to elastic deformations the lattices in the sidebranches are shifted against each other.
Apart from the remarkable feature that the interaction and its origin can be understood fully analytically, we also mention the relation to the similar concept of $\gamma$-surfaces\cite{DickHickelNeu} -- the energy landscape of two half crystals are which are tangentially displaced against each other -- which is essential for the understanding of stacking faults and other defect formation mechanisms.
Here, we obtain a fully analytical prediction of this lateral dependence of this energy landscape for solid-melt-solid layer systems at larger distances.


At this point we also mention that the term interaction must not be interpreted as a mechanical force that leads to an interaction between the grains, but rather a thermodynamic force.
The central difference is that a mechanical attraction e.g.~would move the grains towards each other, so each atom moves.
Here, however, we consider the situation of a melt layer that separates the two grains, and therefore the attraction between the grains would manifest in the solidification of the melt layer.
As a consequence, the gap between the crystal is closed, but without a rigid body motion of the entire grain.
In other words, during the solidification process the number of atoms in the solid phases increases, whereas it would be conserved for a purely mechanical motion.
This aspect is also important from another point of view:
In the consideration of the shifted crystals we have excluded translations in normal direction ($z$ direction) and only investigated motion in the tangential $xy$ plane.
This means that the atomic planes are always aligned in the normal direction and only exhibit a mismatch in the others.
One could also consider the translation in $z$ direction, which would then lead to an additional oscillatory interaction dependence in this direction.
This $\Delta z$ dependence is not related to the exponential decay of the disjoining potential, which appears separately on a larger scale.

Beyond the case of pure grain translation we have also considered grain boundaries with a misorientation.
In this case, a full analytical calculation of the disjoining potential is not possible anymore and numerical investigations are needed \cite{TheNeverendingStory}.
Nevertheless, we have explained that the interaction then stems from higher order terms in the free energy functional, since the longest-range contributions from quadratic terms cancel.
As a result, we find that the disjoining potential decays twice faster then for shifted crystals.
This prediction of the interaction range is confirmed also by numerical simulations and further discussed in Ref.~\onlinecite{TheNeverendingStory}.


\begin{acknowledgments}
R.S. thanks the the DFG Collaborative Research Center 761 {\em Steel -- ab initio} for financial support.
\end{acknowledgments}

\appendix

\section{Linear temperature coupling}
\label{lincoupling}

The difference between the coupling functions (\ref{tilt::eq2}) and (\ref{tilt::eq3}) is that the first is quartic in the amplitude variations in the bulk states and the latter linear.
From this we immediately conclude that in the first case it is not necessary to take the tilt term into account for the solution of the linearized equations in the liquid region, as it is of higher order, and it appears there only effectively through the energy shift as the value of the Hamiltonian.
The situation is different for the linear coupling, and different effects have to be taken into account formally:
(i) The amplitudes in the liquid change since the minimum of the potential energy is shifted away from $\uj=0$ for $T\neq T_M$,
(ii) the amplitudes in the solid change since the minimum of the potential energy is shifted away from $\uj=u_s$ for $T\neq T_M$,
(iii) the thermal tilt gives a contribution to the Hamiltonian in the liquid, which needs to be taken into account up to second order, and 
(iv) the value of the Hamiltonian changes in the solid due to the shift of the solid amplitudes.

At a first glance, one may therefore expect that the results are changed by these effects, and that the long range interaction should depend on the precise choice of the coupling function.
However, here we show that this is not the case up to linear order in $(T-T_M)/T_M$.
For simplicity, we again do not take into account the correction term from the box operator.

Writing $f(\{\uj\}, T) = f_{nonl}(\{\uj\}) + f_T$ in the spirit of Eqs.~(\ref{ae::eq2}) and (\ref{tilt::eq4}), where $f_{nonl}$ therefore contains the free energy density terms to cubic and higher order (leading to the nonlinear terms in the equilibrium conditions), we therefore obtain the equilibrium conditions
\begin{eqnarray}
\frac{\delta F}{\delta \ujs} &=& n_0 k_B T \left( \frac{\uj}{S(q_0)} + \frac{C''(q_0)}{2} (\kvj\cdot\vec{n})^2 \dduj \right) + \nonumber \\
&+& \frac{\partial f_{nonl}(\{\uj\})}{\partial \ujs} + \frac{\partial f_{T}(\{\uj\})}{\partial \ujs} = 0.
\end{eqnarray}
The equilibrium conditions for the liquid between two solid phases are therefore up to first order in the amplitudes
\begin{eqnarray}
&& n_0 k_B T \left( \frac{\uj}{S(q_0)} + \frac{C''(q_0)}{2} (\kvj\cdot\vec{n})^2 \dduj \right) + \nonumber \\
&+& L \frac{T-T_M}{T_M} \frac{1}{N u_s} \frac{\uj}{|\uj|}= 0
\end{eqnarray}
and differ from the previous condition (\ref{single::eq1}) only by the temperature term.
For simplicity, we consider from now only the case of real amplitudes, i.e.~no crystal translation, $\uj/|\uj|=1$.
Then the general solution is
\begin{equation}
\uj = C + c_{j, in} \exp(-\lambda_j z) + c_{j, out} \exp(\lambda_j z)
\end{equation}
where the decay parameters are unchanged and given by Eq.~(\ref{single::eq3}).
The constant term is
\begin{equation}
C = - \frac{S(q_0)}{n_0 k_B T} L \frac{T-T_M}{T_M} \frac{1}{N u_s}.
\end{equation}

In the same way, we can analyze the behavior in the solid phases, where the amplitudes are constant.
One readily finds that the deviation from the previous bulk value $\uj=u_s$ is linear in the temperature deviation $(T-T_M)/T_M$.

The expression for the Hamiltonian up to second order in the amplitudes also now contains the tilt term,
\begin{eqnarray}
H &=& -\frac{n_0 k_B T}{2} \sum_{j=1}^N \Big( \frac{\uj\ujs}{S(q_0)} + \nonumber \\
&=& \frac{C''(q_0)}{2} (\kvj\cdot\vec{n})^2 \duj\dujs \Big) - f_T.
\end{eqnarray}
Inserting the above solution gives up to first order in $(T-T_M)/T_M$ the same expression (\ref{shift::eq7}) as before.

In the solid, the value deviates from the previous expression $H=-\Delta f= L(T-T_M)/T_M$ only by quadratic corrections in $\Delta f$.
Consequently, the long range interaction is unaffected by the choice of the linear instead of a higher order coupling function for the temperature.


\section{The generalized conservation law}
\label{gencons}

As has been demonstrated in Section \ref{InterfaceInteraction} the consideration of the Hamiltonian $H$ as a conserved quantity is a valuable way to understand the interaction of two crystals, which have the same lattice orientation but are shifted against each other.
The limitations were (i) the neglect of the box operator correction and (ii) the restriction to pure translations, which forbids the analysis of grain boundaries, where the crystals are misoriented.

Here we generalize this concept to overcome these restrictions.

For the shifted crystals, the translation leads to a multiplication of the amplitudes by a spatially constant phase factor.
The main difference is that a rotation leads to non-constant phase factors.
In particular, the amplitudes then do not depend only on a single coordinate normal to the grain boundary, but on all spatial coordinates.
Therefore, a proper conservation law should also take into account the directions parallel to the interface.

We consider a ``Lagrangian'' (i.e.~the free energy density in the present context) of the form
\begin{eqnarray} \label{gencons::eq1}
{\cal L} &=& {\cal L}(\{\uj\}, \{\ujs\}, \{\duj\}, \{\dujs\}, \{{\uj}'\}, \{{\ujs}'\}, \nonumber \\
&& \{\dduj\}, \{\ddujs\}, \{{\uj}''\}, \{{\ujs}''\})
\end{eqnarray}
where expressions like $\{\uj\}$ denote the set of all amplitudes $\uj$.
For simplicity, we assume that all fields depend only on two coordinates, which we choose to be normal and tangential to the interface.
This is the case e.g. for tilt grain boundaries, and a twist would require the straightforward inclusion of another tangential dependence.
Derivatives with respect to these directions are denoted by a dot for the normal and a prime for the tangential direction, although this assignment of directions is in principle arbitrary;
however, it will turn out to be a useful choice.
We align our coordinate system such that $x$ is the normal and $y$ the tangential direction.
In contrast to the pure Hamiltonian system in the previous section here also higher order derivatives appear.
It turns out that in our case mixed mode derivatives like $\dujp$ do not appear, and therefore we do not consider them.

We introduce generalized momenta,
\begin{eqnarray}
p^{(j)} := \frac{\partial {\cal L}}{\partial \duj}, &\qquad & q^{(j)} := \frac{\partial {\cal L}}{\partial {\uj}'}, \label{gencons::eq2}\\
r^{(j)} := \frac{\partial {\cal L}}{\partial \dduj}, &\qquad & s^{(j)} := \frac{\partial {\cal L}}{\partial {\uj}''}. \label{gencons::eq3}
\end{eqnarray}
Here, $\uj$ and $\ujs$ are treated as independent variables.
The equilibrium conditions for the amplitudes
\begin{equation}
\frac{\delta}{\delta \uj} \int {\cal L}d\vec{r} = 0
\end{equation}
can be written as
\begin{equation} \label{gencons::eq4}
\frac{\partial {\cal L}}{\partial \uj} - \dot{p}^{(j)} - q^{(j)'} + \ddot{r}^{(j)} + s^{(j)''} = 0.
\end{equation}
We obtain then the following conserved quantity, which means $\dot{H}=0$:
\begin{eqnarray}
H &:=& \int dy \Bigg[ \sum_{j=1}^{N/2} \Big( p^{(j)} \duj + p^{(j)*} \dujs + r^{(j)} \dduj + r^{(j)*} \ddujs \nonumber \\
&&- \dot{r}^{(j)} \duj - \dot{r}^{(j)*} \dujs \Big) -{\cal L} \Bigg]. \label{gencons::eq5}
\end{eqnarray}
Here, $N/2$ is the number of independent density waves (we write the complex conjugate fields explicitly and must not double-count them).
The proof is straightforward:
Performing the normal derivative and application of the equilibrium conditions (\ref{gencons::eq4}) yields after a few algebraic simplifications
\begin{eqnarray*}
\dot{H} &=& \int dy \sum_{i=j}^{N/2} \Bigg[ - \partial_y \left( q^{(j)} \duj + q^{(j)*} \dujs \right) \\
&& - \partial_y \left( s^{(j)} \dot{u}^{(j)'} + s^{(j)*} \dot{u}^{(j)*'} \right) + \partial_y \left( s^{(j)'}\duj\right) \Bigg] \\
&=& 0,
\end{eqnarray*}
where the last step follows from periodicity along the grain boundary (in $y$ direction).

The expressions are written here for a two-dimensional dependence of the fields, which is the case of tilt boundaries in a three-dimensional system.
It is obvious that for more general cases, e.g. twists, the concept can easily be generalized by introduction of a second coordinate in the grain boundary plane.


\end{document}